\documentclass[reprint,amsmath,amssymb,aps,pra]{revtex4-2}

\usepackage{amsthm}
\usepackage{amsfonts}
\usepackage{bbm}
\usepackage{graphicx}
\usepackage{epstopdf}
\usepackage{dcolumn}
\usepackage{bm}
\usepackage{braket}
\usepackage{color}
\usepackage{subfigure}

\usepackage{makecell}
\usepackage[ruled]{algorithm2e}

\begin{document}

\preprint{APS/123-QED}

\title{Detection and evaluation of abnormal user behavior based on quantum generation adversarial network}

\author{Minghua Pan$^{1,\dag}$, Bin Wang$^{1}$, Xiaoling Tao$^{1}$, Shenggen Zheng$^{2}$, Haozhen Situ$^{3,\ddag}$, Lvzhou Li$^{4,\S}$}%
\affiliation{
$^1$ Guangxi Key Laboratory of Cryptography and Information Security, Guilin University of Electronic Technology, Guilin 541004, China \\
$^2$ Peng Cheng Laboratory, Shenzhen 518055, China\\
$^3$ College of Mathematics and Informatics, South China Agricultural University, Guangzhou 510642, China\\
$^4$ Institute of Quantum Computing and Software, School of Computer Science and Engineering, Sun Yat-sen University, Guangzhou 510006, China}
\email{$\dag$panmh@guet.edu.cn;\\
$\ddag$situhaozhen@gmail.com; $\S$lilvzh@mail.sysu.edu.cn
(Corresponding author)}

\date{\today}


\begin{abstract}

Quantum computing holds tremendous potential for processing high-dimensional data, capitalizing on the unique capabilities of superposition and parallelism within quantum states. As we navigate the noisy intermediate-scale quantum (NISQ) era, the exploration of quantum computing applications has emerged as a compelling frontier.
One area of particular interest within the realm of cyberspace security is Behavior Detection and Evaluation (BDE). Notably, the detection and evaluation of internal abnormal behaviors pose significant challenges, given their infrequent occurrence or even their concealed nature amidst vast volumes of normal data.
In this paper, we introduce a novel quantum behavior detection and evaluation algorithm (QBDE) tailored for internal user analysis. The QBDE algorithm comprises a Quantum Generative Adversarial Network (QGAN) in conjunction with a classical neural network for detection and evaluation tasks. The QGAN is built upon a hybrid architecture, encompassing a Quantum Generator ($G_Q$) and a Classical Discriminator ($D_C$). $G_Q$, designed as a parameterized quantum circuit (PQC), collaborates with $D_C$, a classical neural network, to collectively enhance the analysis process.
To address the challenge of imbalanced positive and negative samples, $G_Q$ is employed to generate negative samples. Both $G_Q$ and $D_C$ are optimized through gradient descent techniques. Through extensive simulation tests and quantitative analyses, we substantiate the effectiveness of the QBDE algorithm in detecting and evaluating internal user abnormal behaviors.
Our work not only introduces a novel approach to abnormal behavior detection and evaluation but also pioneers a new application scenario for quantum algorithms. This paradigm shift underscores the promising prospects of quantum computing in tackling complex cybersecurity challenges.

\end{abstract}


\maketitle


\section{Introduction}\label{SecInt}

Quantum computing exploits the quantum parallelism, entanglement, coherence and other properties arising from the superposition of quantum states, and has shown amazing capabilities in processing some computational tasks \cite{Corcoles2019,Zhang2022,Huang2023}. However, current quantum computers are kinds of noisy intermediate-scale quantum (NISQ) computers \cite{Preskill2018}, which have a limited number of qubits and work with noise within a limited coherence time. Exploring possible applications in quantum computing has become an emerging topic in the context of NISQ.

Anomalies are deviations from established patterns within data, and they do not adhere to the predefined norms, as defined by Chandola et al. \cite{Chandola2009}.
 Anomaly detection presents a formidable challenge and has gained paramount importance across numerous research domains, including finance, networking, and health diagnostics.
 Within the realm of network activities, behaviors encompass all actions performed by users, such as login activities, app usage, website visits, and more.
The network's vulnerability to abnormal behaviors, often stemming from the malicious activities of internal users, poses a significant threat to information systems. The intimate knowledge of the system possessed by these users makes the detection of such anomalies particularly challenging.
A noteworthy hurdle in the identification of abnormal behaviors among internal users lies in the pronounced imbalance between negative and positive samples.
In this field, numerous exceptional works have emerged, many of which revolve around training models to capture user behaviors and subsequently assessing whether these behaviors fall within the spectrum of normalcy or exhibit malicious intent \cite{Kim2019,Sharma2020,Singh2020,Malvika2022}.
Conversely, quantum algorithms have begun to find applications in anomaly detection within the realm of physics, notably in scenarios involving quantum data, such as quantum states \cite{Liu2018} and topological phases \cite{Kottmann2021}. A recent breakthrough by Chai et al. \cite{Chai2022QuantumAD} showcased the detection of anomalies within audio samples using a three-qubit quantum spin processor embedded in a diamond.

The Generative Adversarial Networks (GAN) was introduced by Goodfellow et al. in 2014 \cite{Goodfellow2014}. GAN comprises two deep neural networks, namely the generator ($G$) and the discriminator ($D$). Through adversarial training of $G$ and $D$, GAN has the ability to generate synthetic data that closely mimics real data.
GAN has demonstrated remarkable success in modeling complex and high-dimensional distributions of real-world data \cite{Creswell2018GenerativeAN}.
More recently, GAN has found applications in the field of anomaly detection. In 2017, Schlegl et al. proposed AnoGAN (Anomaly GAN), a technique that leverages adversarial training to model normal behavior and calculate anomaly scores for the detection of anomalies \cite{Schlegl2017}.
Numerous enhanced GAN-based methods for anomaly detection have emerged, including EGBAD (Efficient GAN-Based Anomaly Detection) \cite{Zenati2018EfficientGA} and f-AnoGAN (Fast Unsupervised Anomaly Detection with GAN) \cite{Schlegl2019fAnoGAN}.
In 2022, Xia et al. provided a comprehensive review addressing the prominent challenges faced in GAN-based anomaly detection. They also proposed several promising research directions for prediction and analysis in this domain \cite{XIA2022497}.

In 2018, Dallaire-Demers et al. introduced Quantum Generative Adversarial Networks (QGAN) \cite{Demers2018}, expanding the domain of Generative Adversarial Networks (GAN) into the quantum realm. They employed Parameterized Quantum Circuits (PQC) \cite{Benedetti2019PQC} to construct generative adversarial networks and compute gradients, demonstrating the successful training of QGANs. Lloyd et al. \cite{Lloyd2018} further highlighted that QGANs may exhibit an exponential advantage over their classical counterparts, particularly in scenarios involving high-dimensional data samples.
Considering the quantum nature of one or more components, including the generator, discriminator, or data, a diverse array of QGAN algorithm frameworks has emerged \cite{Situ2020,Stein2021,Niu2022,Chaudhary2023}.
In 2021, Herr et al. introduced Variational Quantum-Classical Hybrid Wasserstein GANs (WGANs) \cite{Herr2021}, specifically tailored for anomaly detection within the credit card industry.

To leverage the capabilities of Quantum Generative Adversarial Networks (QGAN) in addressing the challenge of detecting abnormal behaviors among internal users, we propose a variational QGAN, designed using a quantum-classical hybrid architecture within the context of Behavior Detection and Evaluation (BDE).
For simplicity and convenience, we refer to this comprehensive algorithm as QBDE, which stands for "QGAN for Abnormal Behavior Detection and Evaluation based on Internal User Behaviors."
We establish the feasibility and effectiveness of QBDE through a series of simulation experiments conducted using the CERT-R5.2 insider threat test dataset \cite{CERT}. These experiments are executed within the quantum machine learning framework known as PennyLane \cite{Bergholm2018PennyLane}.

The paper is structured as follows.
Sec.\ref{Sec2} is a preliminary about the main processes of GAN and BDE.
Sec.\ref{Sec3} presents our QBDE in detail, including the integration of QGAN and BDE by using parameterized quantum circuits and classical neural networks.
In Sec.\ref{SecResult}, we present the implementation of the QBDE with the insider threat test dataset CERN-R5.2.
Finally, a summary and future works are discussed in Sec.\ref{SecCon}.

\section{Preliminary}\label{Sec2}
In this section, we briefly review the GAN, QGAN and BDE for the abnormal behavior which are the most related techniques of our QBDE.

\subsection{Classical and Quantum Generative Adversarial Networks}\label{SSecGAN}
\begin{figure}\centering
\includegraphics[width=0.42\textwidth]{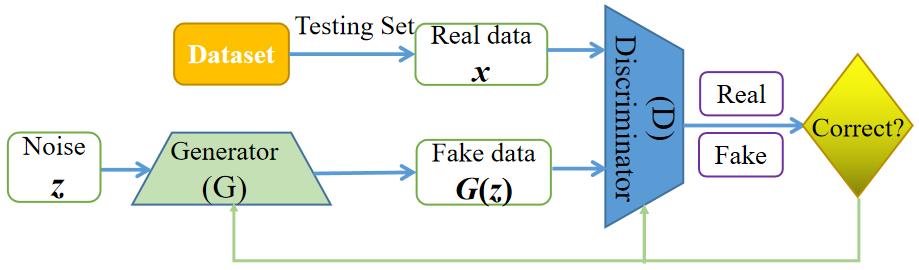}
\caption{\label{fig:GAN}(Color online) The structure of GAN.}
\end{figure}

Generative Adversarial Network (GAN) consists of two neural networks, Generator ($G$) and Discriminator ($D$), which compete against each other like a two-player game \cite{Goodfellow2014}. The structure of GAN is illustrated in Fig. \ref{fig:GAN}. The generator $G$ initiates with a random noise $z$ (a random point in the potential space). Its goal is to learn the probability distribution of samples by training, and subsequently generate the new sample $x= G(z)$, which is similar to the real sample, where $G(\cdot)$ is a function represented by the neural network.
The discriminator $D$ judges the input samples and aims to distinguish the real samples from the generated samples. When the sample $x$ is from the training data set $D_{train}$, the $D$ is trained to assign the data to the ``real" class. For generated samples from the generator, the $D$ is trained to assign them to the ``fake" class.
The $G$ and $D$ compete with each other like game players for optimal performance until reaching "Nash Equilibrium" - a state where $D$ is unable to distinguish between real or generated data.

Let $x$ be a real data subjected to the distribution $P_r$, and $z$ be a noisy sample of the prior probability distribution $P_z$.
The objective function $V (G, D)$ of GAN can be formulated as a min-max optimization problem:
\begin{eqnarray}\label{GANloss}
\min_G\max_D V(G,D)=E_{x\sim{P_r(x)}}[\log D(x)]\nonumber\\
+E_{x\sim{P_z(x)}}[\log (1-D(G(z)))].
\end{eqnarray}
where $D(G(z))$ is the probability that $D$ classifies fake data as real.

In practice, $G$ and $D$ are trained iteratively by minimizing their respective loss functions. Considering binary classification, let $y=1$ for real data and $y=0$ for fake data.
The loss function of $G$ is
\begin{eqnarray}\label{Gloss}
L_G=[(1-y)][\log (1-D(G(z)))].
\end{eqnarray}
The loss function of $D$ is
\begin{eqnarray}\label{Dloss}
L_D=-[y\log D(x)+(1-y)\log(1-D(G(z)))].
\end{eqnarray}

By training the discriminator $D$ against the generator $G$, the ability of $G$ to generate realistic samples is constantly improved, and $D$ is also improved in correctly identifying real and generated samples. In the ideal case, when the game reaches Nash Equilibrium, the distribution of the data generated by the generator fits that of the real data.

Quantum generative adversarial network (QGAN) is a generalized version of classical GAN using quantum properties. For the generator, discriminator and data, it is considered as a QGAN if one or more of them is quantum. In current NISQ, a large number of QGANs adopt the quantum-classical hybrid architecture that just $G$ or $D$ is quantum. For the quantum part, they usually apply PQCs to construct the quantum circuit of $G$ or $D$. We will introduce our QBDE as the specific case in Sec.\ref{Sec3}.

\subsection{The abnormal behavior detection and evaluation based on user behaviors}\label{SSecBDE}

\begin{figure}\centering
\includegraphics[width=0.48\textwidth]{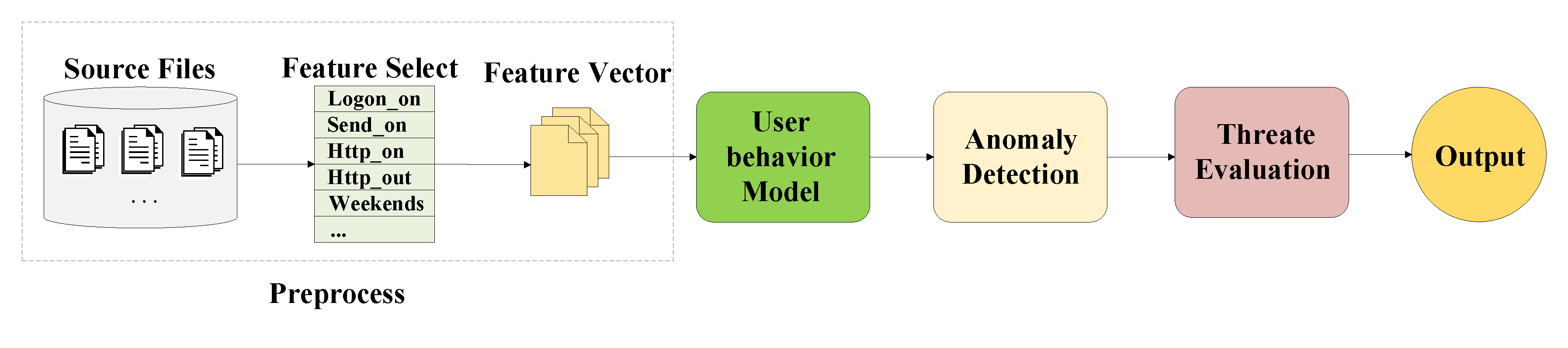}
\caption{\label{fig:BDE}(Color online) The process of abnormal behavior detection and evaluation.}
\end{figure}

The process of abnormal behavior detection and evaluation  is shown in FIG. \ref{fig:BDE}, which generally includes the following processes (More details see Ref.\cite{Creswell2018GenerativeAN,Sharma2020}). It can be divided into three parts: preprocessing, behavior modeling, detection and evaluation.

\textbf{Preprocessing.}
First, the original user datasets are selected from different multiple files. Then, behavior features are extracted. Considering the features are varied for different users, the data needs to be divided into separate datasets for each user. Further, the features are normalized into feature vectors.

\textbf{Behavior modeling.}
A user behavior model is necessary to evaluate users' behaviors. Ref.\cite{Sharma2020} proposed a user behaviors model training based on normal behavior sequence with GAN. During the training stage, the reconstruction is carried out in the output to minimize the reconstruction error \cite{Malhotra2016}. In the test stage, samples including normal or abnormal are fed to the network. For the unknown abnormal data, the network will produce a high reconstruction error. Thus, the unknown user behavior is correctly judged.

\textbf{Behavior detection and evaluation (BDE).}
To detect abnormal behaviors, test data is fed into the trained normal behavior model. Then, the threat degree of the detected threatening behavior is evaluated by a behavior detection and evaluation (BDE) network. In order to evaluate the security of a user's behavior, the behavior score $d(x)$ and the abnormal threshold $Th_d$ are required.

For the testing data $x$ and the generated data $G(z)$, let $R_d$ and $R_n$ be the reconstruction errors before and after passing through the network of BDE. We have
\begin{eqnarray}\label{Rd}
R_d=\parallel x-G(z)\parallel_1,
\end{eqnarray}
and
\begin{eqnarray}\label{Rn}
R_n=\parallel f_n(x)-f_n(G(z))\parallel_1,
\end{eqnarray}
where $f_n(\cdot)$ represents the function of the BDE network, $\parallel \alpha \parallel_1$ is the $l_1$ norm of $\alpha$.

The behavior score $d(x)$ is defined as
\begin{eqnarray}\label{Dbs}
d(x)=(1-\lambda)R_d+ \lambda R_n,
\end{eqnarray}
where $\lambda$ represents the weight.

The abnormal threshold depends on the specific task. During the detection stage, a behavior $X_t(x)$ can be classified as either 'Normal' or 'Abnormal' based on the behavior score $d(x)$ and the threshold $Th_d$ as
\begin{eqnarray}\label{THd}
X_t(x)=\left\{
\begin{aligned}
&Normal, &&d(x)\leq Th_d, x\in D_{test},\\
&Abnormal, &&d(x)> Th_d, x\in D_{test}.
\end{aligned}
\right.
\end{eqnarray}

The main purpose of abnormal detection and evaluation is to analyze the threat level of user behavior, so as to defend and protect the networks and systems. The evaluation function $f(d(x))$ and the threat threshold $Th_f$ of the abnormal user behavior are used to achieve the above aim. Then, an abnormal behavior is divided into two threat levels, $Low\_threat$ and $High\_threat$, according to $f(d(x))$ and $Th_f$ in the following way
\begin{eqnarray}\label{B}
f(d(x))=\left\{
\begin{aligned}
&Low\_threat, &&d(x)\leq Th_f,\\
&Higt\_threat, &&d(x)> Th_f.
\end{aligned}
\right.
\end{eqnarray}
$Low\_threat$ indicates no malicious behavior or a lower frequency abnormal operations, while $High \_threat$ indicates malicious behavior or higher frequent abnormal operations.

\section{User abnormal behavior detection and evaluation based on QGAN}\label{Sec3}

\begin{figure}\centering
\includegraphics[width=0.45\textwidth]{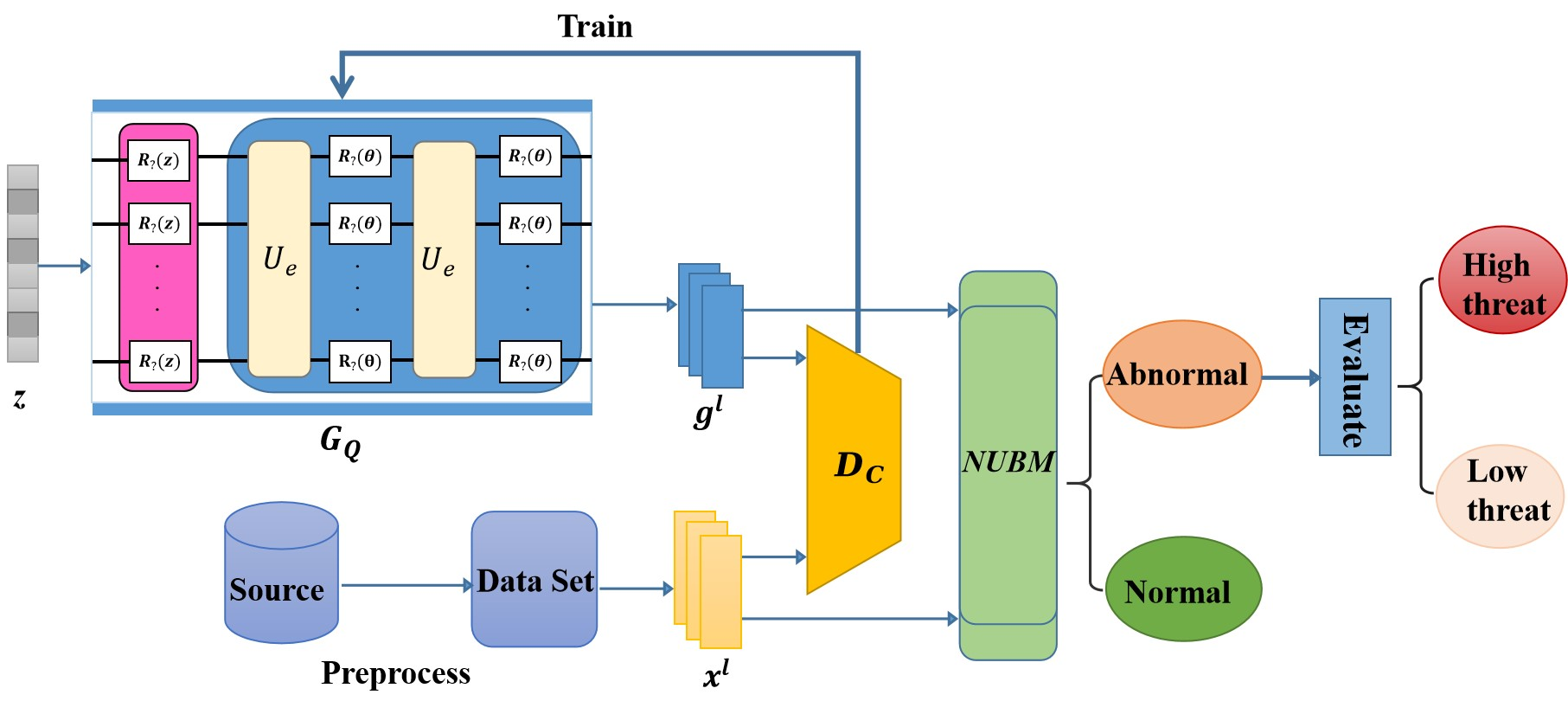}
\caption{\label{fig:fQBDE}(Color online) The framework of QBDE. Here, $G_Q$, $D_C$, $UNBM$ are the quantum Generator, the classical Discriminator and the user normal behavior model, respectively.}
\end{figure}

Due to the data generation ability of GAN and the superiority of quantum-classical hybrid architecture, we propose a quantum BDE algorithm, \textbf{QBDE}. The QBDE detects and evaluates user abnormal behavior based on a quantum generative adversarial network (QGAN). The framework of QBDE is shown in Fig.\ref{fig:fQBDE}. Similar to the classical abnormal behavior detection and evaluation model, it includes three modules: data preprocessing, the construction of the normal user behavior model (NUBM), and the behavior detection and evaluation (BDE). Specifically, the construction of NUBM is implemented by QGAN, which consists a quantum Generator $G_Q$ and a classical Discriminator $D_C$. Hence, our focus will be on the QGAN applied in the NUBM stage as well as the BDE in the following subsections.

\subsection{Quantum generative adversarial network for constructing user behavior model}\label{SSecQGAN}
In the anomaly detection, the samples of abnormal behavior are generally lesser than the normal samples. Especially, for the internal abnormal behaviors, it occurs much less often and are even covered by a large amount of normal data. In other words, the proportion of positive and negative samples is extremely imbalanced. Therefore, we select the normal user behaviors to form the training set, apply QGAN to generate negative samples, and then train the network to implement the construction of NUBM. Considering the limited resources of current quantum systems, the QGAN in the QBDE adopts a hybrid quantum-classical architecture, where the generator is a PQC and the discriminator is a classical neural network.

\subsubsection{Quantum Generator} \label{SSSecQG}
\begin{figure}\centering
\subfigure[The structure of $G_Q$] {\label{fig:subfig:GQ}\includegraphics[width=0.4\textwidth]{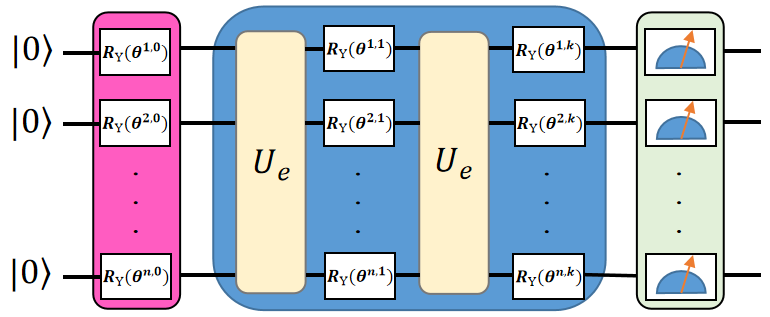}}\quad
\subfigure[The entangled gate $U_e$] {\label{fig:subfig:Ue}\includegraphics[width=0.3\textwidth]{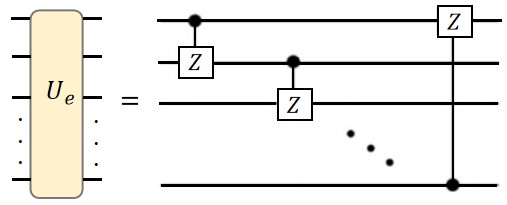}}\quad
\caption{\label{fig:PQC}(Color online) The structure of $G_Q$ and its entangled gate.}
\end{figure}

The quantum generator $G_Q$ of QBDE adopts PQC architecture \cite{Benedetti2019PQC}, which consists of a series of single parameterized quantum gates and controlled quantum gates. Considering that the data for detecting anomaly user behavior is discrete, the special PQC architecture in $G_Q$ of QBDE is shown in Fig.\ref{fig:subfig:GQ}, which is proposed in Ref. \cite{Zoufal2019}. Each layer is composed of a series of single rotation Pauli-Y gates $R_Y(\theta^{i,j})$ and entangled gates $U_e$, where $\theta^{i,j}$ represents the rotation angle of the $i$th qubit in layer $j$, and $U_e$ is composed of multiple controlled gates $Z$ as shown in Fig.\ref{fig:subfig:Ue}. The rotation gates and entangled gates are executed alternately. Assume the system consists of $n$ qubits, and let $K$ denote the depth of a quantum circuit. The $G_Q$ is trained to convert a given input state $\ket{\psi_{in}}$ into the output state
\begin{eqnarray}
\ket{g_\theta}=G_\theta\ket{\psi_{in}}=\sum_{j=0}^{2^n-1}\sqrt{p_\theta^j}\ket{j},
\end{eqnarray}
where $p_\theta^j$ is the probability of state $\ket{j}$, $G_\theta$ represents the parameterized quantum circuit for $G_Q$ with the parameter $\theta$.
To be specific, the input state is
$\ket{\psi_{in}}=R_Y(\theta^0){\ket{0...00}}$,
and the $G_\theta$ can be expressed as
\begin{eqnarray}
G_\theta=R_Y(\theta^K) U_e...R_Y(\theta^2) U_e R_Y(\theta^1)U_e,
\end{eqnarray}
where $R_Y(\theta^j)=R_Y(\theta^{1,j})\otimes R_Y(\theta^{2,j})\otimes...\otimes R_Y(\theta^{n,j})$ is the rotation Pauli-Y gates in the $j$-th layer.

\subsubsection{Classical Discriminator}\label{SSSecCD}
\begin{figure}\centering
\includegraphics[width=0.36\textwidth]{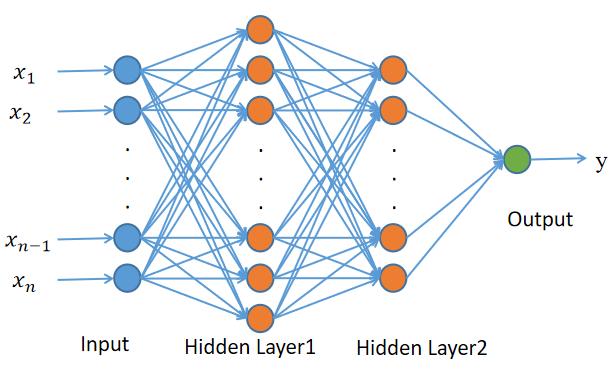}
\caption{\label{fig:DC}(Color online) The network of the classical Discriminator $D_C$.}
\end{figure}

The power of $G_Q$ is limited due to the restricted number of qubits and circuit depth in current quantum systems. Under this circumstance,
it is not suitable to choose a complex network in the discriminator, so as to ensure that the $D_C$ does not overwhelm the $G_Q$ \cite{Zoufal2019}. Therefore, we employ a fully connected neural network as the classical discriminator $D_C$ of QBDE, as depicted in Fig. \ref{fig:DC}. The  structure of $D_C$ consists of two hidden layers composed of fully connected neurons and one neuron output layer.

\begin{table*}
\caption{Files and features of the selected data}\label{Tab1}
\begin{tabular}	{ c c c c }
\hline
File Name&Description about the data&Features\\ \hline
login.csv& System login record&$login\_on, loginoff\_on, login\_out, loginoff\_out, weekend$\\
http.csv& Http access record&$http\_on, http\_out$\\ 
device.cvs& Mobile Device usage record&$connect\_on, disconnect\_on, connect\_out, disconnect\_out, size$\\
email.csv& Sending and receiving of mails&$send\_on, send\_out$\\ 
file.csv& File operation record&$file\_on, file\_off$\\ \hline
\end{tabular}
\end{table*}

\subsubsection{Training and Optimization}\label{SSSecOb}
The objective function of GAN is a min-max optimization problem, as stated in Subsection \ref{SSecGAN}. The goal of $G_Q$ is to learn the probability distribution of samples by training and generate newly generated samples that resemble real ones.
The $D_C$ judges the input samples and aims to distinguish between the real samples and the generated samples.
Considering the practical application, the batch processing technology is adopted. Assuming that the batch size is $m$, $x^l$ and $g^l$ are the training and generated samples. The loss function of the $G_Q$ for QBDE becomes
\begin{eqnarray}\label{QTDEGloss}
L_G(\theta,\phi)=\frac{1}{m}\sum_{l=1}^m[\log D_\phi(g^l)].
\end{eqnarray}
The loss function of the $D_C$ is
\begin{eqnarray}\label{QTDEDloss}
L_D(\theta,\phi)=-\frac{1}{m}\sum_{l=1}^m[\log D_\phi(g^l)+\log(1- D_\phi(g^l))].
\end{eqnarray}
$L_G$ and $L_D$ are alternately optimized by the parameters $\theta$ of $G_Q$ and the parameter $\phi$ of $D_C$. We optimize them by PennyLane \cite{Bergholm2018PennyLane}, which is a software framework for differentiable programming of quantum computers. It computes the gradient of a variational quantum circuit in a way that is compatible with classical techniques. 

\subsection{The abnormal behavior detection and evaluation}\label{SSecIUADE}
\begin{figure}\centering
\includegraphics[width=0.3\textwidth]{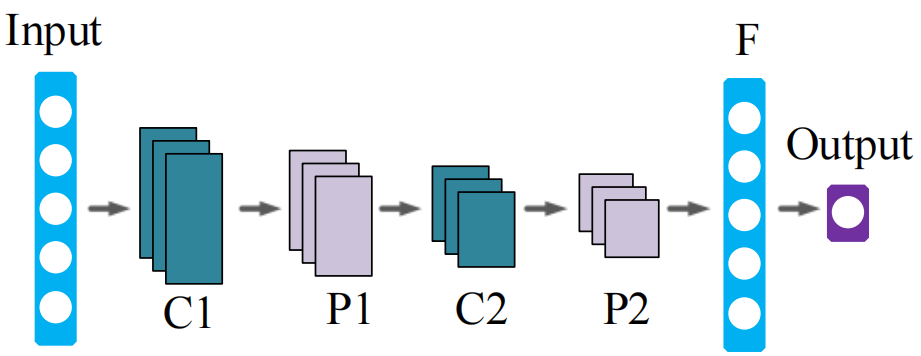}
\caption{\label{fig:Det}(Color online) The network of BDE for internal user.}
\end{figure}

The anomaly behavior detection and evaluation model (BDE) utilizes a two-classes convolutional neural network, as shown in Fig. \ref{fig:Det}. This network comprises composed of two convolution layers, two maximum pooling layers and the final output layer. The output layer consists of only one neuron with the Sigmoid activation function.

In order to validate the effectiveness of the designed algorithm, we evaluate the accuracy of the classification of the tested behaviors and the loss of BDE. The accuracy is expressed as
\begin{eqnarray}\label{Acc}
Accuracy =\frac{TP+TP}{TP+TN+FP+FN},
\end{eqnarray}
where TP, TN represent the ratio of true positive (negative) samples, and FP, FN are the ratio of false positive (negative) samples predicted wrong by the network.

The BDE network is optimized by minimizing its cost function, which is represented by the cross entropy loss: 
\begin{eqnarray}\label{Biloss}
J(w,b)=-\frac{1}{m}\sum_{i=1}^m[y_i\log(\hat{y_i})+(1-y_i)\log(1-\hat{y_i})],
\end{eqnarray}
where $y_i$ is the real label and $\hat{y_i}$ is the output value of $x_i$, respectively. Here, $\hat{y_i}=\omega x_i +b$ is determined by the BDE network with parameters $\omega$ and $b$.

Since the training set only contains normal samples, we take the maximum reconstruction error of all training data as the abnormal threshold $Th_d$, i.e
\begin{eqnarray}\label{Thd2}
Th_d=\max(d(x_1),d(x_2),...,d(x_i)), x_i\in D_{train}.
\end{eqnarray}

In general, when the user takes malicious behavior or performs many abnormal operations, the behavior score $d(x)$ is more than twice of $Th_d$. Therefore, we take the threat threshold
\begin{eqnarray}\label{Thf2}
Th_f=2Th_d.
\end{eqnarray}

\section{Experiments and result analysis}\label{SecResult}

\subsection{Experimental environment and data}
The experiments were performed in Win10 with 1T, the Intel $i5-9500$ processor and 16G memory. The programming platforms and software used included Python 3.8.12, PyCharm Community 2020.3, the machine learning library TensorFlow and PennyLane.

The data is from the well-known insider threat test data set CERT-IT R5.2 \cite{CERT}. It includes the simulated attack behaviors such as system destruction, information theft and internal fraud carried by malicious internal users, as well as a large amount of normal behavior data. It consists of multiple files, which contain various log data of employee behaviors in the organization. 
Data files are processed in parallel according to user names. Then, for each individual user, behavior data is collected by day as features. 
Here, we select the data from $5$ specific files and extract $16$ behavior features as shown in Table \ref{Tab1}. In the features column, `$\_on$', `$\_out$' represent the records that occurred during the working time and out of working time, respectively.

For user behaviors, some values of features are much large than others. Hence, we normalize the values of features to $[0,1]$ in the following way
\begin{eqnarray}\label{Norm}
x_{i,j}'=\frac{x_{i,j}-\min(x_j)}{\max(x_j)-\min(x_j)},x_{i,j}\in {x_i}, j\in[1,16].
\end{eqnarray}
where $x_{i,j}$ is the value of the row $i$ (the $i$th day) and column $j$ (the $j$th feature) in any matrix $X$ within the dataset, and $\min(x_j)$ and $\max(x_j)$ are the minimum and maximum of the $j$-th feature, respectively.

For the input of $G_Q$, we only need $4$ qubits for $16$ features of user behaviors, and initialize them with $\ket{0000}$ at first. The input state $\ket\psi_{in}$ will be initialized by adjusting $\theta^{i,0}$ of quantum rotation gate $R_Y(\theta^{i,0})$ as
\begin{eqnarray}\label{Input}
\ket{\psi_{in}}=R_Y(\theta^{i,0})\ket{0000}, i\in[1,4].
\end{eqnarray}
where $i$ is the $i$th qubit.

\begin{algorithm}[p]
	\caption{QGAN for user behavior detection and evaluation, QBDE}
	\label{AlgQBDE}
	\LinesNumbered
	\KwIn{Data set $D$, the initial quantum state $\ket{0000}$.}
	\KwOut{Normal, Low threat, High threat.}
  Preprocess the original user data set, normalize the data into a feature vector, and divide them into $D_{train}$ and $D_{test}$.\\
  Training QGAN with PennyLane:\\
    \For{training iterations}{
  \begin{itemize}
    \item Prepare the initial quantum state $\ket{\psi_{in}}$ by adjusting $R_Y(\theta)$.
   \item Generate $\{\ket{g^l}\}$ using PQC, and obtain $\{g^l\}$ by measurement.
   \item Sample $\{x^l\}$ from $D_{train}$.
  \end{itemize}
    \For{epoch size}{
      \begin{itemize}
     \item  Update the $D_C$ by descending its stochastic gradient $\bigtriangledown_{\theta\phi}L_D$ according to Eq.(\ref{QTDEDloss}).\\
    \item  Update the $G_Q$ by descending its stochastic gradient $\bigtriangledown_{\theta\phi}L_G$ according to Eq.(\ref{QTDEGloss}).
      \end{itemize}
    }
  }
  Fed the generating data and training sets into the NUBM and train it with cost function $J(\omega,b)$.\\
  Detection and evaluation:\\
  Calculate the behavior score $d(x)$, and compare it with $Th_d$:\\
    \uIf{$d(x) < Th_d$}{
    Normal
  }
  \ElseIf{$Th_d < d(x) <Th_f$}{
    Low thread
  }
  \Else{
    High thread
  }
\end{algorithm}

\subsection{Algorithm Implementation of QBDE}\label{SSecAl}
Based on the theory of QBDE we proposed in the previous Sec.\ref{Sec3}, the overall procedure of QBED is summarized with pseudo-code in Algorithm \ref{AlgQBDE}.
The input includes the data set $D$ selected from CERT-IT R5.2 \cite{CERT} and the quantum state $\ket{0000}$.
The QBED is first implemented by QGAN which is trained by PennyLane with the gradient descent, then a new behavior is detected and evaluated by BDE.

\begin{figure}\centering
\includegraphics[width=0.4\textwidth]{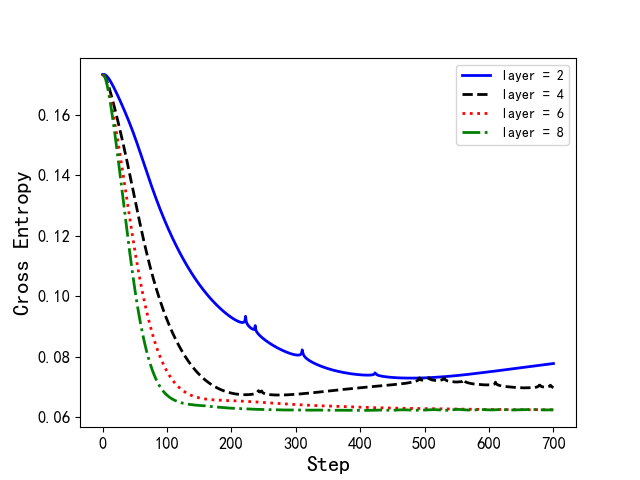}
\caption{\label{fig:entropy}(Color online) The $cross\_entropy$ vs. the epoch for different layers in QGAN.}
\end{figure}

\begin{figure}\centering
\subfigure[$K=2$]{\label{fig:subfig:adgloss}\includegraphics[width=0.22\textwidth]{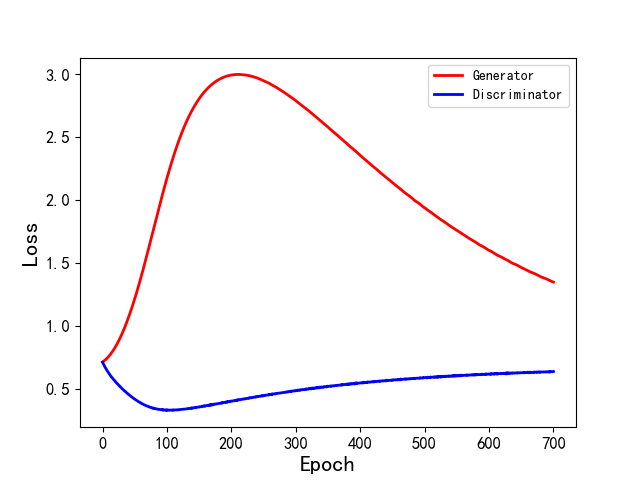}}\quad
\subfigure[$K=4$]{\label{fig:subfig:bdgloss}\includegraphics[width=0.22\textwidth]{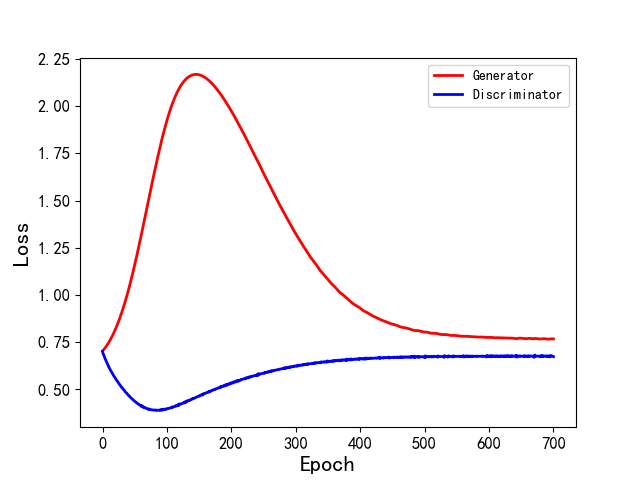}}\quad\\
\subfigure[$K=6$]{\label{fig:subfig:cdgloss}\includegraphics[width=0.22\textwidth]{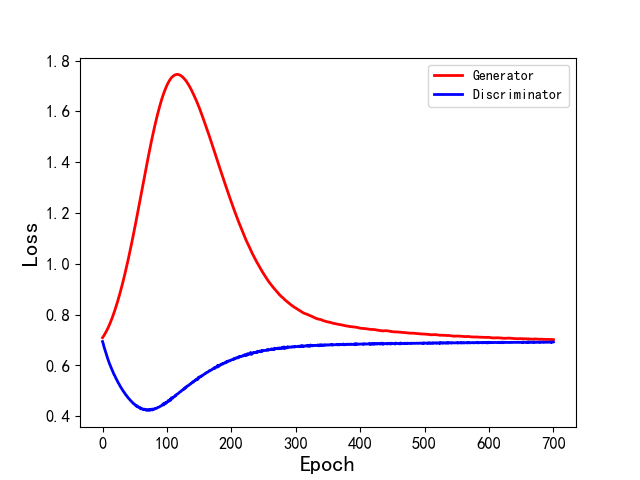}}\quad
\subfigure[$K=8$]{\label{fig:subfig:ddgloss}\includegraphics[width=0.22\textwidth]{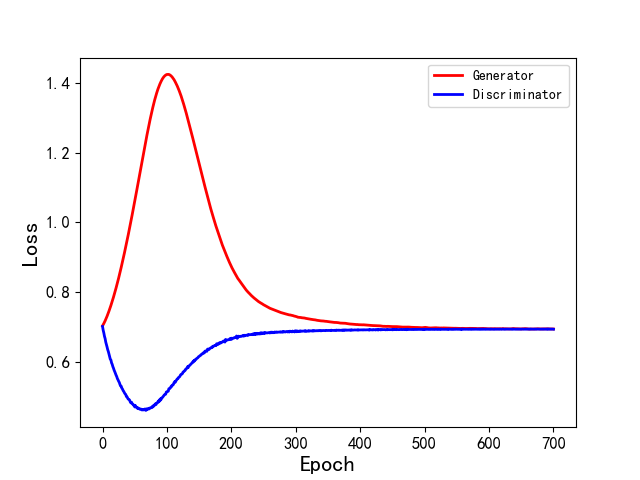}}\quad
\caption{\label{fig:dgloss}(Color online) The losses of $G_Q$ and $D_C$ vs. the epoch for different layers in QGAN.}
\end{figure}

\begin{figure}\centering
\subfigure[$K=2$]{\label{fig:subfig:score22}\includegraphics[width=0.22\textwidth]{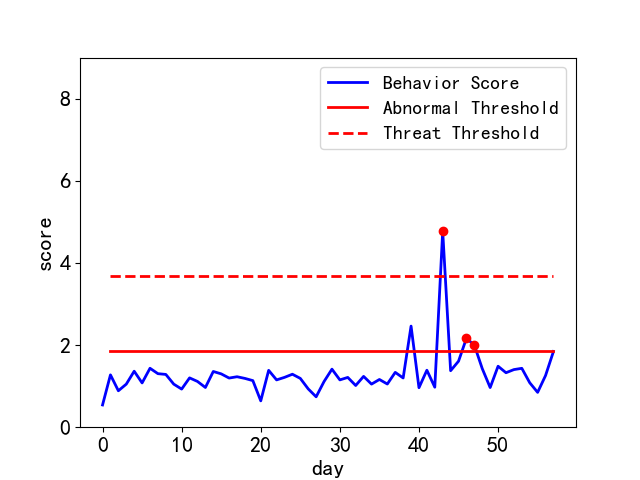}}\quad
\subfigure[$K=4$]{\label{fig:subfig:score24}\includegraphics[width=0.22\textwidth]{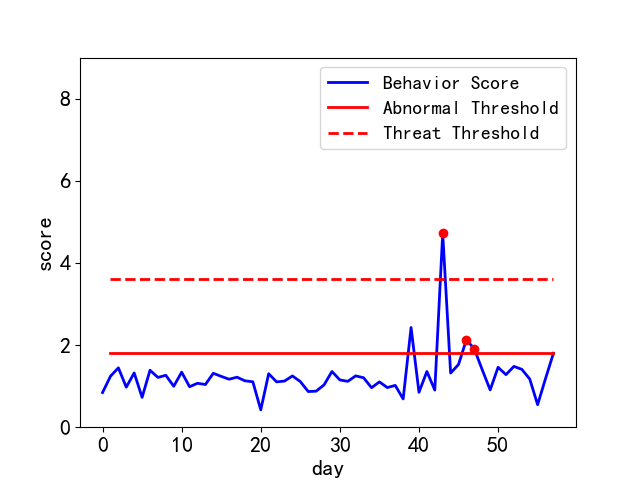}}\quad\\
\subfigure[$K=6$]{\label{fig:subfig:score26}\includegraphics[width=0.22\textwidth]{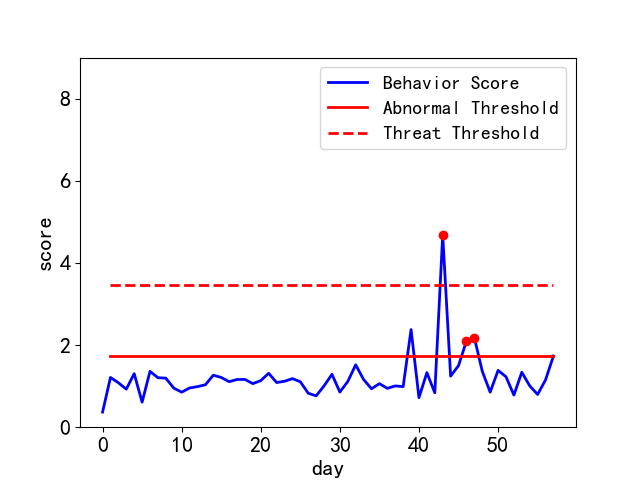}}\quad
\subfigure[$K=8$]{\label{fig:subfig:score28}\includegraphics[width=0.22\textwidth]{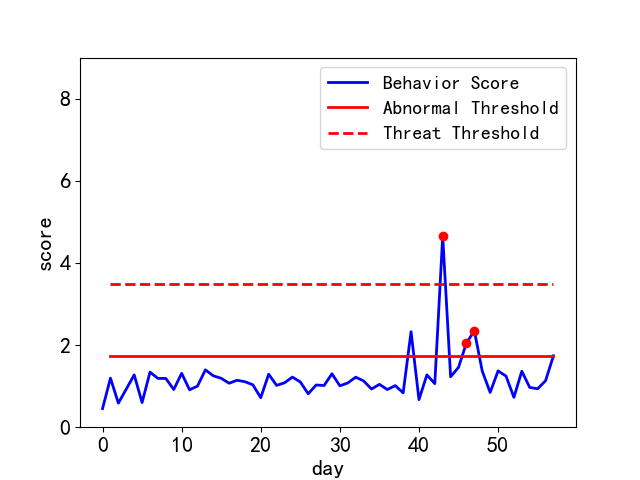}}\quad
\caption{\label{fig:score}(Color online) The scores of threat $d(x)$ vs. the days for different layers in QGAN. The red dots represent the real abnormal behaviors that happen on that day.}
\end{figure}

\begin{figure*}\centering
\subfigure[User1]{\label{fig:subfig:U18loss}\includegraphics[width=0.3\textwidth]{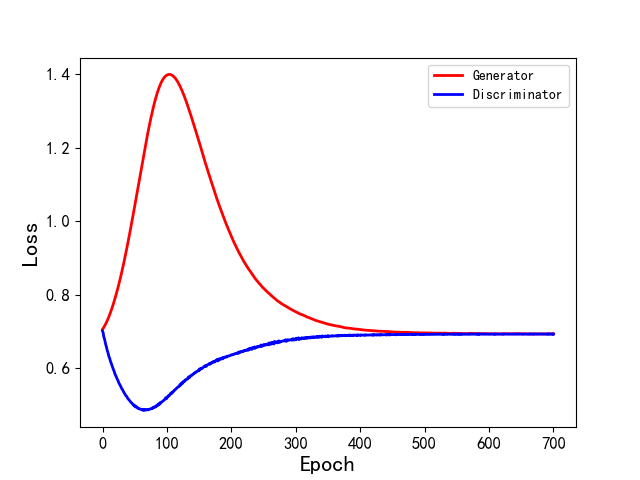}}\quad
\subfigure[User2]{\label{fig:subfig:U28loss}\includegraphics[width=0.3\textwidth]{28loss.png}}\quad
\subfigure[User3]{\label{fig:subfig:U38loss}\includegraphics[width=0.3\textwidth]{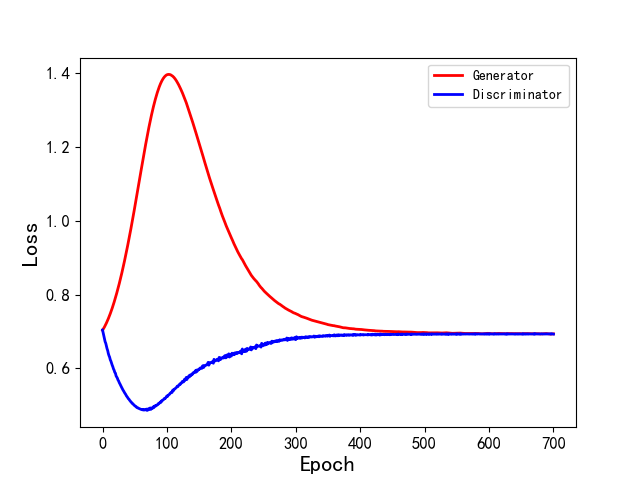}}\quad
\caption{\label{fig:Ugdloss}(Color online) The losses of $G_Q$ and $D_C$ vs. the epoch for different users.}
\end{figure*}

\begin{figure*}\centering
\subfigure[User1]{\label{fig:subfig:Uscore18}\includegraphics[width=0.3\textwidth]{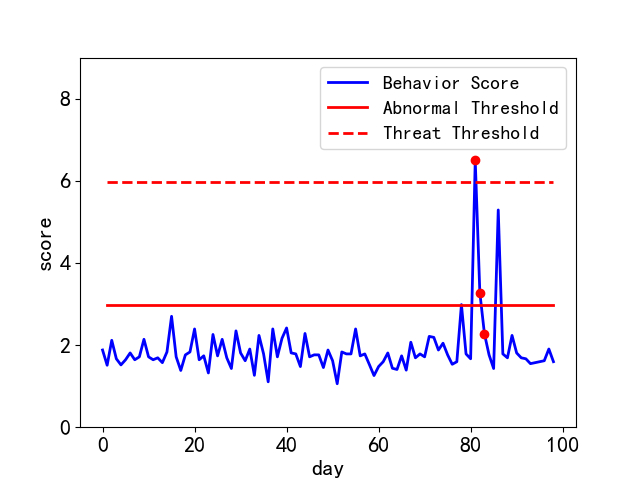}}\quad
\subfigure[User2]{\label{fig:subfig:Uscore28}\includegraphics[width=0.3\textwidth]{score28.png}}\quad
\subfigure[User3]{\label{fig:subfig:Uscore38}\includegraphics[width=0.3\textwidth]{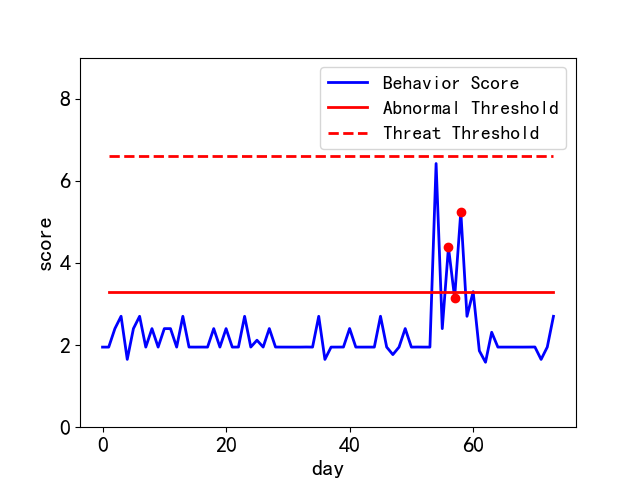}}\quad
\caption{\label{fig:Uscore}(Color online) The score of threat $d(x)$ vs. the days for different users. The red dots represent the real abnormal behaviors happen in that day.}
\end{figure*}

\subsection{Experimental results and analysis} \label{SSecTest}

\subsubsection{QBDE with different layers of PQC} \label{SSSecLayer}
The depth of the neural network plays a crucial role in its training process. In the case of QGAN, where $G_Q$ adopts the PQC architecture, the performance of $G_Q$ is influenced by the depth (i.e. the number of layers) of PQC. Considering that the depth is limited in NISQ, we investigate the performance of $G_Q$ with different layers $K=2,4,6,8$. In terms of the average cross-entropy and the loss functions, the performance of QGAN is shown in Fig.\ref{fig:entropy} and Fig.\ref{fig:dgloss}.

The observation in Fig. \ref{fig:entropy} reveals that with an increasing number of epochs, the cross entropy of $G_Q$ decreases quickly at first, and then slows down after a certain epoch, indicating that the optimization of $G_Q$ tends to converge.
The convergence is inadequate for $K=2,4$, but highly satisfactory for $K=6,8$.
Moreover, it is clear that the cross-entropy of $G_Q$ decreases more rapidly  as the number of layers $K$ increases. In other words, deeper depths result in faster and superior convergence of $G_Q$.

In Fig. \ref{fig:dgloss}, the loss $L_G$ increases first and then decreases rapidly. On the contrary, the loss $L_D$ initially decreases and then increases. Eventually, both $L_G$ and $L_D$ converge to similar values and remain relatively constant with the epoch, indicating that means the samples generated by $G_Q$ are already equivalent to the real samples. Additionally, the more layers there are, the faster $L_G$ and $L_D$ tend to converge.

The results of the detection, in terms of the behavioral score $d(x)$ of the test data, are depicted in Fig. \ref{fig:score}. The days of abnormal behaviors occurring in the real world were marked with red dots. We can  observe that with different layers of QGAN, the abnormal behaviors have been successfully detected and evaluated with the behaviors scores. It is noted that several normal behaviors have higher scores than the $Th_d$ which may be classified as abnormal. Anyhow, the accuracy of the QBDE is $98.28\%$.

\subsubsection{QBDE for different users}
Now, we consider three different users. User1 has 300 days of data, with 200 days are used for training and 100 days for testing; User2 has 160 days of data, with 100 days for training and 60 days for testing; User3 has 175 days of data, with 100 days for training and 75 days for testing.
Here, the number of layers of PQC in QGAN is $K=8$.
The losses of the $G_Q$ and $D_C$ are depicted in Fig.\ref{fig:Ugdloss}, while the behavior scores $d(x)$ of test data are illustrated in Fig.\ref{fig:Uscore}.
From these figures, we can observe that $L_G$ and $L_D$ tend to have the similar values for different users, respectively. Meanwhile, almost all behaviors can be successfully detected and evaluated. For these three users, we obtain the accuracies of the QBDEs are $97.98\%,\ 98.28\%,\ 97.30\%$.

Based on the above discussions, we can conclude that QBDE with QGAN can efficiently generate fake samples to construct the normal user behavior model, which can be further applied for the internal user abnormal behavior detection and evaluation.

\section{Conclusion}\label{SecCon}

In this paper, we introduced QBDE, a Quantum Generative Adversarial Network founded upon a quantum-classical hybrid architecture. QBDE was developed to address the challenge of detecting and evaluating abnormal behaviors among internal users. The quantum generator within QBDE played a pivotal role in generating negative samples, effectively mitigating the imbalance issue between positive and negative samples—a common challenge when dealing with limited abnormal behavior data. Furthermore, both the quantum generator and classical discriminator were optimized using the PennyLane framework.
Our experiments demonstrated the feasibility and efficacy of QBDE when applied to the CERT-R5.2 insider threat test dataset.
However, there remains ample room for improvement. The current QBDE implementation utilizes a limited number of qubits and shallow Parameterized Quantum Circuits (PQC), which constrains the potential of $G_Q$. Additionally, these limitations extend to the utilization of advanced and complex neural networks for $D_C$.
This designed algorithm not only paves the way for new applications in quantum artificial intelligence but also introduces a novel approach to abnormal behavior detection. In light of this, there exist numerous avenues for future research in exploring further applications for quantum algorithms.

\section*{Acknowledgments}
M. Pan was supported by the National Natural Science Foundation of China (Grant No. 62361021), Guangxi Science and Technology Program (No.GuiKeAD21075020), Guangxi Key Laboratory of Cryptography and Information Security (No.202133).
B. Wang was supported by the Innovation Project of GUET Graduate Education (No. 2022YCXS068).
S.Z acknowledge supports in part from the Major Key Project of PCL, Innovation Program for Quantum Science and Technology (2021ZD0302900). Li was supported by the National Natural Science Foundation of China (Grant No.  62272492).

\nocite{*}
\bibliography{QBDE}

\begin{thebibliography}{40}%
\makeatletter
\providecommand \@ifxundefined [1]{%
 \@ifx{#1\undefined}
}%
\providecommand \@ifnum [1]{%
 \ifnum #1\expandafter \@firstoftwo
 \else \expandafter \@secondoftwo
 \fi
}%
\providecommand \@ifx [1]{%
 \ifx #1\expandafter \@firstoftwo
 \else \expandafter \@secondoftwo
 \fi
}%
\providecommand \natexlab [1]{#1}%
\providecommand \enquote  [1]{``#1''}%
\providecommand \bibnamefont  [1]{#1}%
\providecommand \bibfnamefont [1]{#1}%
\providecommand \citenamefont [1]{#1}%
\providecommand \href@noop [0]{\@secondoftwo}%
\providecommand \href [0]{\begingroup \@sanitize@url \@href}%
\providecommand \@href[1]{\@@startlink{#1}\@@href}%
\providecommand \@@href[1]{\endgroup#1\@@endlink}%
\providecommand \@sanitize@url [0]{\catcode `\\12\catcode `\$12\catcode
  `\&12\catcode `\#12\catcode `\^12\catcode `\_12\catcode `\%12\relax}%
\providecommand \@@startlink[1]{}%
\providecommand \@@endlink[0]{}%
\providecommand \url  [0]{\begingroup\@sanitize@url \@url }%
\providecommand \@url [1]{\endgroup\@href {#1}{\urlprefix }}%
\providecommand \urlprefix  [0]{URL }%
\providecommand \Eprint [0]{\href }%
\providecommand \doibase [0]{https://doi.org/}%
\providecommand \selectlanguage [0]{\@gobble}%
\providecommand \bibinfo  [0]{\@secondoftwo}%
\providecommand \bibfield  [0]{\@secondoftwo}%
\providecommand \translation [1]{[#1]}%
\providecommand \BibitemOpen [0]{}%
\providecommand \bibitemStop [0]{}%
\providecommand \bibitemNoStop [0]{.\EOS\space}%
\providecommand \EOS [0]{\spacefactor3000\relax}%
\providecommand \BibitemShut  [1]{\csname bibitem#1\endcsname}%
\let\auto@bib@innerbib\@empty
\bibitem [{\citenamefont {C{\'o}rcoles}\ \emph {et~al.}(2019)\citenamefont
  {C{\'o}rcoles}, \citenamefont {Kandala}, \citenamefont {Javadi-Abhari},
  \citenamefont {McClure}, \citenamefont {Cross}, \citenamefont {Temme},
  \citenamefont {Nation}, \citenamefont {Steffen},\ and\ \citenamefont
  {Gambetta}}]{Corcoles2019}%
  \BibitemOpen
  \bibfield  {author} {\bibinfo {author} {\bibfnamefont {A.~D.}\ \bibnamefont
  {C{\'o}rcoles}}, \bibinfo {author} {\bibfnamefont {A.}~\bibnamefont
  {Kandala}}, \bibinfo {author} {\bibfnamefont {A.}~\bibnamefont
  {Javadi-Abhari}}, \bibinfo {author} {\bibfnamefont {D.~T.}\ \bibnamefont
  {McClure}}, \bibinfo {author} {\bibfnamefont {A.~W.}\ \bibnamefont {Cross}},
  \bibinfo {author} {\bibfnamefont {K.}~\bibnamefont {Temme}}, \bibinfo
  {author} {\bibfnamefont {P.~D.}\ \bibnamefont {Nation}}, \bibinfo {author}
  {\bibfnamefont {M.}~\bibnamefont {Steffen}},\ and\ \bibinfo {author}
  {\bibfnamefont {J.~M.}\ \bibnamefont {Gambetta}},\ }\bibfield  {title}
  {\bibinfo {title} {Challenges and opportunities of near-term quantum
  computing systems},\ }\href@noop {} {\bibfield  {journal} {\bibinfo
  {journal} {Proceedings of the IEEE}\ }\textbf {\bibinfo {volume} {108}},\
  \bibinfo {pages} {1338} (\bibinfo {year} {2019})}\BibitemShut {NoStop}%
\bibitem [{\citenamefont {Zhang}\ and\ \citenamefont {Li}(2022)}]{Zhang2022}%
  \BibitemOpen
  \bibfield  {author} {\bibinfo {author} {\bibfnamefont {S.}~\bibnamefont
  {Zhang}}\ and\ \bibinfo {author} {\bibfnamefont {L.}~\bibnamefont {Li}},\
  }\bibfield  {title} {\bibinfo {title} {A brief introduction to quantum
  algorithms},\ }\href@noop {} {\bibfield  {journal} {\bibinfo  {journal} {CCF
  Transactions on High Performance Computing}\ }\textbf {\bibinfo {volume}
  {4}},\ \bibinfo {pages} {53} (\bibinfo {year} {2022})}\BibitemShut {NoStop}%
\bibitem [{\citenamefont {Huang}\ \emph {et~al.}(2023)\citenamefont {Huang},
  \citenamefont {Xu}, \citenamefont {Guo}, \citenamefont {Tian}, \citenamefont
  {Wei}, \citenamefont {Sun}, \citenamefont {Bao},\ and\ \citenamefont
  {Long}}]{Huang2023}%
  \BibitemOpen
  \bibfield  {author} {\bibinfo {author} {\bibfnamefont {H.-L.}\ \bibnamefont
  {Huang}}, \bibinfo {author} {\bibfnamefont {X.-Y.}\ \bibnamefont {Xu}},
  \bibinfo {author} {\bibfnamefont {C.}~\bibnamefont {Guo}}, \bibinfo {author}
  {\bibfnamefont {G.}~\bibnamefont {Tian}}, \bibinfo {author} {\bibfnamefont
  {S.-J.}\ \bibnamefont {Wei}}, \bibinfo {author} {\bibfnamefont
  {X.}~\bibnamefont {Sun}}, \bibinfo {author} {\bibfnamefont {W.-S.}\
  \bibnamefont {Bao}},\ and\ \bibinfo {author} {\bibfnamefont {G.-L.}\
  \bibnamefont {Long}},\ }\bibfield  {title} {\bibinfo {title} {Near-term
  quantum computing techniques: Variational quantum algorithms, error
  mitigation, circuit compilation, benchmarking and classical simulation},\
  }\href@noop {} {\bibfield  {journal} {\bibinfo  {journal} {SCIENCE
  CHINA-PHYSICS MECHANICS \& ASTRONOMY}\ }\textbf {\bibinfo {volume} {66}}
  (\bibinfo {year} {2023})}\BibitemShut {NoStop}%
\bibitem [{\citenamefont {Preskill}(2018)}]{Preskill2018}%
  \BibitemOpen
  \bibfield  {author} {\bibinfo {author} {\bibfnamefont {J.}~\bibnamefont
  {Preskill}},\ }\bibfield  {title} {\bibinfo {title} {Quantum {C}omputing in
  the {NISQ} era and beyond},\ }\href@noop {} {\bibfield  {journal} {\bibinfo
  {journal} {{Quantum}}\ }\textbf {\bibinfo {volume} {2}},\ \bibinfo {pages}
  {79} (\bibinfo {year} {2018})}\BibitemShut {NoStop}%
\bibitem [{\citenamefont {Chandola}\ \emph {et~al.}(2009)\citenamefont
  {Chandola}, \citenamefont {Banerjee},\ and\ \citenamefont
  {Kumar}}]{Chandola2009}%
  \BibitemOpen
  \bibfield  {author} {\bibinfo {author} {\bibfnamefont {V.}~\bibnamefont
  {Chandola}}, \bibinfo {author} {\bibfnamefont {A.}~\bibnamefont {Banerjee}},\
  and\ \bibinfo {author} {\bibfnamefont {V.}~\bibnamefont {Kumar}},\ }\bibfield
   {title} {\bibinfo {title} {Anomaly detection: A survey},\ }\href@noop {}
  {\bibfield  {journal} {\bibinfo  {journal} {ACM Comput. Surv.}\ }\textbf
  {\bibinfo {volume} {41}},\ \bibinfo {pages} {15:1} (\bibinfo {year}
  {2009})}\BibitemShut {NoStop}%
\bibitem [{\citenamefont {Kim}\ \emph {et~al.}(2019)\citenamefont {Kim},
  \citenamefont {Park}, \citenamefont {Kim}, \citenamefont {Cho},\ and\
  \citenamefont {Kang}}]{Kim2019}%
  \BibitemOpen
  \bibfield  {author} {\bibinfo {author} {\bibfnamefont {J.}~\bibnamefont
  {Kim}}, \bibinfo {author} {\bibfnamefont {M.}~\bibnamefont {Park}}, \bibinfo
  {author} {\bibfnamefont {H.}~\bibnamefont {Kim}}, \bibinfo {author}
  {\bibfnamefont {S.}~\bibnamefont {Cho}},\ and\ \bibinfo {author}
  {\bibfnamefont {P.}~\bibnamefont {Kang}},\ }\bibfield  {title} {\bibinfo
  {title} {Insider threat detection based on user behavior modeling and anomaly
  detection algorithms},\ }\href@noop {} {\bibfield  {journal} {\bibinfo
  {journal} {Applied Sciences}\ }\textbf {\bibinfo {volume} {9}},\ \bibinfo
  {pages} {4018} (\bibinfo {year} {2019})}\BibitemShut {NoStop}%
\bibitem [{\citenamefont {Sharma}\ \emph {et~al.}(2020)\citenamefont {Sharma},
  \citenamefont {Pokharel},\ and\ \citenamefont {Joshi}}]{Sharma2020}%
  \BibitemOpen
  \bibfield  {author} {\bibinfo {author} {\bibfnamefont {B.}~\bibnamefont
  {Sharma}}, \bibinfo {author} {\bibfnamefont {P.}~\bibnamefont {Pokharel}},\
  and\ \bibinfo {author} {\bibfnamefont {B.}~\bibnamefont {Joshi}},\ }\bibfield
   {title} {\bibinfo {title} {User behavior analytics for anomaly detection
  using lstm autoencoder - insider threat detection},\ }\href@noop {}
  {\bibfield  {journal} {\bibinfo  {journal} {Proceedings of the 11th
  International Conference on Advances in Information Technology}\ } (\bibinfo
  {year} {2020})}\BibitemShut {NoStop}%
\bibitem [{\citenamefont {Singh}\ \emph {et~al.}(2020)\citenamefont {Singh},
  \citenamefont {Mehtre},\ and\ \citenamefont {Sangeetha}}]{Singh2020}%
  \BibitemOpen
  \bibfield  {author} {\bibinfo {author} {\bibfnamefont {M.}~\bibnamefont
  {Singh}}, \bibinfo {author} {\bibfnamefont {B.~M.}\ \bibnamefont {Mehtre}},\
  and\ \bibinfo {author} {\bibfnamefont {S.}~\bibnamefont {Sangeetha}},\
  }\bibfield  {title} {\bibinfo {title} {Insider threat detection based on user
  behaviour analysis},\ }in\ \href@noop {} {\emph {\bibinfo {booktitle}
  {Machine Learning, Image Processing, Network Security and Data Sciences}}}\
  (\bibinfo  {publisher} {Springer Singapore},\ \bibinfo {address}
  {Singapore},\ \bibinfo {year} {2020})\ pp.\ \bibinfo {pages}
  {559--574}\BibitemShut {NoStop}%
\bibitem [{\citenamefont {Singh}\ \emph {et~al.}(2022)\citenamefont {Singh},
  \citenamefont {Mehtre},\ and\ \citenamefont {Sangeetha}}]{Malvika2022}%
  \BibitemOpen
  \bibfield  {author} {\bibinfo {author} {\bibfnamefont {M.}~\bibnamefont
  {Singh}}, \bibinfo {author} {\bibfnamefont {B.~M.}\ \bibnamefont {Mehtre}},\
  and\ \bibinfo {author} {\bibfnamefont {S.}~\bibnamefont {Sangeetha}},\
  }\bibfield  {title} {\bibinfo {title} {User behavior based insider threat
  detection using a multi fuzzy classifier},\ }\href@noop {} {\bibfield
  {journal} {\bibinfo  {journal} {Multimedia Tools and Applications}\ }\textbf
  {\bibinfo {volume} {81}},\ \bibinfo {pages} {22953} (\bibinfo {year}
  {2022})}\BibitemShut {NoStop}%
\bibitem [{\citenamefont {Liu}\ and\ \citenamefont
  {Rebentrost}(2018)}]{Liu2018}%
  \BibitemOpen
  \bibfield  {author} {\bibinfo {author} {\bibfnamefont {N.}~\bibnamefont
  {Liu}}\ and\ \bibinfo {author} {\bibfnamefont {P.}~\bibnamefont
  {Rebentrost}},\ }\bibfield  {title} {\bibinfo {title} {Quantum machine
  learning for quantum anomaly detection},\ }\href@noop {} {\bibfield
  {journal} {\bibinfo  {journal} {Phys. Rev. A}\ }\textbf {\bibinfo {volume}
  {97}},\ \bibinfo {pages} {042315} (\bibinfo {year} {2018})}\BibitemShut
  {NoStop}%
\bibitem [{\citenamefont {Kottmann}\ \emph {et~al.}(2021)\citenamefont
  {Kottmann}, \citenamefont {Metz}, \citenamefont {Fraxanet},\ and\
  \citenamefont {Baldelli}}]{Kottmann2021}%
  \BibitemOpen
  \bibfield  {author} {\bibinfo {author} {\bibfnamefont {K.}~\bibnamefont
  {Kottmann}}, \bibinfo {author} {\bibfnamefont {F.}~\bibnamefont {Metz}},
  \bibinfo {author} {\bibfnamefont {J.}~\bibnamefont {Fraxanet}},\ and\
  \bibinfo {author} {\bibfnamefont {N.}~\bibnamefont {Baldelli}},\ }\bibfield
  {title} {\bibinfo {title} {Variational quantum anomaly detection:
  Unsupervised mapping of phase diagrams on a physical quantum computer},\
  }\href@noop {} {\bibfield  {journal} {\bibinfo  {journal} {Phys. Rev.
  Research}\ }\textbf {\bibinfo {volume} {3}},\ \bibinfo {pages} {043184}
  (\bibinfo {year} {2021})}\BibitemShut {NoStop}%
\bibitem [{\citenamefont {Chai}\ \emph {et~al.}(2022)\citenamefont {Chai},
  \citenamefont {Liu}, \citenamefont {Wang}, \citenamefont {Guo}, \citenamefont
  {Shi}, \citenamefont {Li}, \citenamefont {Wang},\ and\ \citenamefont
  {Du}}]{Chai2022QuantumAD}%
  \BibitemOpen
  \bibfield  {author} {\bibinfo {author} {\bibfnamefont {Z.}~\bibnamefont
  {Chai}}, \bibinfo {author} {\bibfnamefont {Y.}~\bibnamefont {Liu}}, \bibinfo
  {author} {\bibfnamefont {M.}~\bibnamefont {Wang}}, \bibinfo {author}
  {\bibfnamefont {Y.}~\bibnamefont {Guo}}, \bibinfo {author} {\bibfnamefont
  {F.}~\bibnamefont {Shi}}, \bibinfo {author} {\bibfnamefont {Z.}~\bibnamefont
  {Li}}, \bibinfo {author} {\bibfnamefont {Y.}~\bibnamefont {Wang}},\ and\
  \bibinfo {author} {\bibfnamefont {J.}~\bibnamefont {Du}},\ }\bibfield
  {title} {\bibinfo {title} {Quantum anomaly detection of audio samples with a
  spin processor in diamond},\ }\href@noop {} {\bibfield  {journal} {\bibinfo
  {journal} {ArXiv}\ }\textbf {\bibinfo {volume} {abs/2201.10263}} (\bibinfo
  {year} {2022})}\BibitemShut {NoStop}%
\bibitem [{\citenamefont {Goodfellow}\ \emph {et~al.}(2014)\citenamefont
  {Goodfellow}, \citenamefont {Pouget-Abadie}, \citenamefont {Mirza},
  \citenamefont {Xu}, \citenamefont {Warde-Farley}, \citenamefont {Ozair},
  \citenamefont {Courville},\ and\ \citenamefont {Bengio}}]{Goodfellow2014}%
  \BibitemOpen
  \bibfield  {author} {\bibinfo {author} {\bibfnamefont {I.~J.}\ \bibnamefont
  {Goodfellow}}, \bibinfo {author} {\bibfnamefont {J.}~\bibnamefont
  {Pouget-Abadie}}, \bibinfo {author} {\bibfnamefont {M.}~\bibnamefont
  {Mirza}}, \bibinfo {author} {\bibfnamefont {B.}~\bibnamefont {Xu}}, \bibinfo
  {author} {\bibfnamefont {D.}~\bibnamefont {Warde-Farley}}, \bibinfo {author}
  {\bibfnamefont {S.}~\bibnamefont {Ozair}}, \bibinfo {author} {\bibfnamefont
  {A.}~\bibnamefont {Courville}},\ and\ \bibinfo {author} {\bibfnamefont
  {Y.}~\bibnamefont {Bengio}},\ }\bibfield  {title} {\bibinfo {title}
  {Generative adversarial nets},\ }in\ \href@noop {} {\emph {\bibinfo
  {booktitle} {Proceedings of the 27th International Conference on Neural
  Information Processing Systems - Volume 2}}},\ \bibinfo {series and number}
  {NIPS'14}\ (\bibinfo  {publisher} {MIT Press},\ \bibinfo {address}
  {Cambridge, MA, USA},\ \bibinfo {year} {2014})\ pp.\ \bibinfo {pages}
  {2672--2680}\BibitemShut {NoStop}%
\bibitem [{\citenamefont {Creswell}\ \emph {et~al.}(2018)\citenamefont
  {Creswell}, \citenamefont {White}, \citenamefont {Dumoulin}, \citenamefont
  {Arulkumaran}, \citenamefont {Sengupta},\ and\ \citenamefont
  {Bharath}}]{Creswell2018GenerativeAN}%
  \BibitemOpen
  \bibfield  {author} {\bibinfo {author} {\bibfnamefont {A.}~\bibnamefont
  {Creswell}}, \bibinfo {author} {\bibfnamefont {T.}~\bibnamefont {White}},
  \bibinfo {author} {\bibfnamefont {V.}~\bibnamefont {Dumoulin}}, \bibinfo
  {author} {\bibfnamefont {K.}~\bibnamefont {Arulkumaran}}, \bibinfo {author}
  {\bibfnamefont {B.}~\bibnamefont {Sengupta}},\ and\ \bibinfo {author}
  {\bibfnamefont {A.~A.}\ \bibnamefont {Bharath}},\ }\bibfield  {title}
  {\bibinfo {title} {Generative adversarial networks: An overview},\
  }\href@noop {} {\bibfield  {journal} {\bibinfo  {journal} {IEEE Signal
  Processing Magazine}\ }\textbf {\bibinfo {volume} {35}},\ \bibinfo {pages}
  {53} (\bibinfo {year} {2018})}\BibitemShut {NoStop}%
\bibitem [{\citenamefont {Schlegl}\ \emph {et~al.}(2017)\citenamefont
  {Schlegl}, \citenamefont {Seeb{\"o}ck}, \citenamefont {Waldstein},
  \citenamefont {Schmidt-Erfurth},\ and\ \citenamefont {Langs}}]{Schlegl2017}%
  \BibitemOpen
  \bibfield  {author} {\bibinfo {author} {\bibfnamefont {T.}~\bibnamefont
  {Schlegl}}, \bibinfo {author} {\bibfnamefont {P.}~\bibnamefont
  {Seeb{\"o}ck}}, \bibinfo {author} {\bibfnamefont {S.~M.}\ \bibnamefont
  {Waldstein}}, \bibinfo {author} {\bibfnamefont {U.~M.}\ \bibnamefont
  {Schmidt-Erfurth}},\ and\ \bibinfo {author} {\bibfnamefont {G.}~\bibnamefont
  {Langs}},\ }\bibfield  {title} {\bibinfo {title} {Unsupervised anomaly
  detection with generative adversarial networks to guide marker discovery},\
  }in\ \href@noop {} {\emph {\bibinfo {booktitle} {IPMI}}}\ (\bibinfo {year}
  {2017})\BibitemShut {NoStop}%
\bibitem [{\citenamefont {Zenati}\ \emph {et~al.}(2018)\citenamefont {Zenati},
  \citenamefont {Foo}, \citenamefont {Lecouat}, \citenamefont {Manek},\ and\
  \citenamefont {Chandrasekhar}}]{Zenati2018EfficientGA}%
  \BibitemOpen
  \bibfield  {author} {\bibinfo {author} {\bibfnamefont {H.}~\bibnamefont
  {Zenati}}, \bibinfo {author} {\bibfnamefont {C.-S.}\ \bibnamefont {Foo}},
  \bibinfo {author} {\bibfnamefont {B.}~\bibnamefont {Lecouat}}, \bibinfo
  {author} {\bibfnamefont {G.}~\bibnamefont {Manek}},\ and\ \bibinfo {author}
  {\bibfnamefont {V.~R.}\ \bibnamefont {Chandrasekhar}},\ }\bibfield  {title}
  {\bibinfo {title} {Efficient gan-based anomaly detection},\ }\href@noop {}
  {\bibfield  {journal} {\bibinfo  {journal} {ArXiv}\ }\textbf {\bibinfo
  {volume} {abs/1802.06222}} (\bibinfo {year} {2018})}\BibitemShut {NoStop}%
\bibitem [{\citenamefont {Schlegl}\ \emph {et~al.}(2019)\citenamefont
  {Schlegl}, \citenamefont {Seeb{\"o}ck}, \citenamefont {Waldstein},
  \citenamefont {Langs},\ and\ \citenamefont
  {Schmidt-Erfurth}}]{Schlegl2019fAnoGAN}%
  \BibitemOpen
  \bibfield  {author} {\bibinfo {author} {\bibfnamefont {T.}~\bibnamefont
  {Schlegl}}, \bibinfo {author} {\bibfnamefont {P.}~\bibnamefont
  {Seeb{\"o}ck}}, \bibinfo {author} {\bibfnamefont {S.~M.}\ \bibnamefont
  {Waldstein}}, \bibinfo {author} {\bibfnamefont {G.}~\bibnamefont {Langs}},\
  and\ \bibinfo {author} {\bibfnamefont {U.~M.}\ \bibnamefont
  {Schmidt-Erfurth}},\ }\bibfield  {title} {\bibinfo {title} {f-anogan: Fast
  unsupervised anomaly detection with generative adversarial networks},\
  }\href@noop {} {\bibfield  {journal} {\bibinfo  {journal} {Medical Image
  Analysis}\ }\textbf {\bibinfo {volume} {54}},\ \bibinfo {pages} {30¨C44}
  (\bibinfo {year} {2019})}\BibitemShut {NoStop}%
\bibitem [{\citenamefont {Xia}\ \emph {et~al.}(2022)\citenamefont {Xia},
  \citenamefont {Pan}, \citenamefont {Li}, \citenamefont {He}, \citenamefont
  {Ma}, \citenamefont {Zhang},\ and\ \citenamefont {Ding}}]{XIA2022497}%
  \BibitemOpen
  \bibfield  {author} {\bibinfo {author} {\bibfnamefont {X.}~\bibnamefont
  {Xia}}, \bibinfo {author} {\bibfnamefont {X.}~\bibnamefont {Pan}}, \bibinfo
  {author} {\bibfnamefont {N.}~\bibnamefont {Li}}, \bibinfo {author}
  {\bibfnamefont {X.}~\bibnamefont {He}}, \bibinfo {author} {\bibfnamefont
  {L.}~\bibnamefont {Ma}}, \bibinfo {author} {\bibfnamefont {X.}~\bibnamefont
  {Zhang}},\ and\ \bibinfo {author} {\bibfnamefont {N.}~\bibnamefont {Ding}},\
  }\bibfield  {title} {\bibinfo {title} {Gan-based anomaly detection: A
  review},\ }\href@noop {} {\bibfield  {journal} {\bibinfo  {journal}
  {Neurocomputing}\ }\textbf {\bibinfo {volume} {493}},\ \bibinfo {pages} {497}
  (\bibinfo {year} {2022})}\BibitemShut {NoStop}%
\bibitem [{\citenamefont {Dallaire-Demers}\ and\ \citenamefont
  {Killoran}(2018)}]{Demers2018}%
  \BibitemOpen
  \bibfield  {author} {\bibinfo {author} {\bibfnamefont {P.-L.}\ \bibnamefont
  {Dallaire-Demers}}\ and\ \bibinfo {author} {\bibfnamefont {N.}~\bibnamefont
  {Killoran}},\ }\bibfield  {title} {\bibinfo {title} {Quantum generative
  adversarial networks},\ }\href@noop {} {\bibfield  {journal} {\bibinfo
  {journal} {Phys. Rev. A}\ }\textbf {\bibinfo {volume} {98}},\ \bibinfo
  {pages} {012324} (\bibinfo {year} {2018})}\BibitemShut {NoStop}%
\bibitem [{\citenamefont {Benedetti}\ \emph
  {et~al.}(2019{\natexlab{a}})\citenamefont {Benedetti}, \citenamefont {Lloyd},
  \citenamefont {Sack},\ and\ \citenamefont {Fiorentini}}]{Benedetti2019PQC}%
  \BibitemOpen
  \bibfield  {author} {\bibinfo {author} {\bibfnamefont {M.}~\bibnamefont
  {Benedetti}}, \bibinfo {author} {\bibfnamefont {E.}~\bibnamefont {Lloyd}},
  \bibinfo {author} {\bibfnamefont {S.~H.}\ \bibnamefont {Sack}},\ and\
  \bibinfo {author} {\bibfnamefont {M.}~\bibnamefont {Fiorentini}},\ }\bibfield
   {title} {\bibinfo {title} {Parameterized quantum circuits as machine
  learning models},\ }\href@noop {} {\bibfield  {journal} {\bibinfo  {journal}
  {Quantum Science and Technology}\ }\textbf {\bibinfo {volume} {4}} (\bibinfo
  {year} {2019}{\natexlab{a}})}\BibitemShut {NoStop}%
\bibitem [{\citenamefont {Lloyd}\ and\ \citenamefont
  {Weedbrook}(2018)}]{Lloyd2018}%
  \BibitemOpen
  \bibfield  {author} {\bibinfo {author} {\bibfnamefont {S.}~\bibnamefont
  {Lloyd}}\ and\ \bibinfo {author} {\bibfnamefont {C.}~\bibnamefont
  {Weedbrook}},\ }\bibfield  {title} {\bibinfo {title} {Quantum generative
  adversarial learning},\ }\href@noop {} {\bibfield  {journal} {\bibinfo
  {journal} {Phys. Rev. Lett.}\ }\textbf {\bibinfo {volume} {121}},\ \bibinfo
  {pages} {040502} (\bibinfo {year} {2018})}\BibitemShut {NoStop}%
\bibitem [{\citenamefont {Situ}\ \emph {et~al.}(2020)\citenamefont {Situ},
  \citenamefont {He}, \citenamefont {Wang}, \citenamefont {Li},\ and\
  \citenamefont {Zheng}}]{Situ2020}%
  \BibitemOpen
  \bibfield  {author} {\bibinfo {author} {\bibfnamefont {H.}~\bibnamefont
  {Situ}}, \bibinfo {author} {\bibfnamefont {Z.}~\bibnamefont {He}}, \bibinfo
  {author} {\bibfnamefont {Y.}~\bibnamefont {Wang}}, \bibinfo {author}
  {\bibfnamefont {L.}~\bibnamefont {Li}},\ and\ \bibinfo {author}
  {\bibfnamefont {S.}~\bibnamefont {Zheng}},\ }\bibfield  {title} {\bibinfo
  {title} {Quantum generative adversarial network for generating discrete
  distribution},\ }\href@noop {} {\bibfield  {journal} {\bibinfo  {journal}
  {Information Sciences}\ }\textbf {\bibinfo {volume} {538}},\ \bibinfo {pages}
  {193} (\bibinfo {year} {2020})}\BibitemShut {NoStop}%
\bibitem [{\citenamefont {Stein}\ \emph {et~al.}(2021)\citenamefont {Stein},
  \citenamefont {Baheri}, \citenamefont {Chen}, \citenamefont {Mao},
  \citenamefont {Guan}, \citenamefont {Li}, \citenamefont {Fang},\ and\
  \citenamefont {Xu}}]{Stein2021}%
  \BibitemOpen
  \bibfield  {author} {\bibinfo {author} {\bibfnamefont {S.~A.}\ \bibnamefont
  {Stein}}, \bibinfo {author} {\bibfnamefont {B.}~\bibnamefont {Baheri}},
  \bibinfo {author} {\bibfnamefont {D.}~\bibnamefont {Chen}}, \bibinfo {author}
  {\bibfnamefont {Y.}~\bibnamefont {Mao}}, \bibinfo {author} {\bibfnamefont
  {Q.}~\bibnamefont {Guan}}, \bibinfo {author} {\bibfnamefont {A.}~\bibnamefont
  {Li}}, \bibinfo {author} {\bibfnamefont {B.}~\bibnamefont {Fang}},\ and\
  \bibinfo {author} {\bibfnamefont {S.}~\bibnamefont {Xu}},\ }\bibfield
  {title} {\bibinfo {title} {Qugan: A quantum state fidelity based generative
  adversarial network},\ }in\ \href@noop {} {\emph {\bibinfo {booktitle} {2021
  IEEE International Conference on Quantum Computing and Engineering (QCE)}}}\
  (\bibinfo {year} {2021})\ pp.\ \bibinfo {pages} {71--81}\BibitemShut
  {NoStop}%
\bibitem [{\citenamefont {Niu}\ \emph {et~al.}(2022)\citenamefont {Niu},
  \citenamefont {Zlokapa}, \citenamefont {Broughton}, \citenamefont {Boixo},
  \citenamefont {Mohseni}, \citenamefont {Smelyanskyi},\ and\ \citenamefont
  {Neven}}]{Niu2022}%
  \BibitemOpen
  \bibfield  {author} {\bibinfo {author} {\bibfnamefont {M.~Y.}\ \bibnamefont
  {Niu}}, \bibinfo {author} {\bibfnamefont {A.}~\bibnamefont {Zlokapa}},
  \bibinfo {author} {\bibfnamefont {M.}~\bibnamefont {Broughton}}, \bibinfo
  {author} {\bibfnamefont {S.}~\bibnamefont {Boixo}}, \bibinfo {author}
  {\bibfnamefont {M.}~\bibnamefont {Mohseni}}, \bibinfo {author} {\bibfnamefont
  {V.}~\bibnamefont {Smelyanskyi}},\ and\ \bibinfo {author} {\bibfnamefont
  {H.}~\bibnamefont {Neven}},\ }\bibfield  {title} {\bibinfo {title}
  {Entangling quantum generative adversarial networks},\ }\href@noop {}
  {\bibfield  {journal} {\bibinfo  {journal} {Phys. Rev. Lett.}\ }\textbf
  {\bibinfo {volume} {128}},\ \bibinfo {pages} {220505} (\bibinfo {year}
  {2022})}\BibitemShut {NoStop}%
\bibitem [{\citenamefont {Chaudhary}\ \emph {et~al.}(2023)\citenamefont
  {Chaudhary}, \citenamefont {Huembeli}, \citenamefont {MacCormack},
  \citenamefont {Patti}, \citenamefont {Kossaifi},\ and\ \citenamefont
  {Galda}}]{Chaudhary2023}%
  \BibitemOpen
  \bibfield  {author} {\bibinfo {author} {\bibfnamefont {S.}~\bibnamefont
  {Chaudhary}}, \bibinfo {author} {\bibfnamefont {P.}~\bibnamefont {Huembeli}},
  \bibinfo {author} {\bibfnamefont {I.}~\bibnamefont {MacCormack}}, \bibinfo
  {author} {\bibfnamefont {T.~L.}\ \bibnamefont {Patti}}, \bibinfo {author}
  {\bibfnamefont {J.}~\bibnamefont {Kossaifi}},\ and\ \bibinfo {author}
  {\bibfnamefont {A.}~\bibnamefont {Galda}},\ }\bibfield  {title} {\bibinfo
  {title} {Towards a scalable discrete quantum generative adversarial neural
  network},\ }\href {https://dx.doi.org/10.1088/2058-9565/acc4e4} {\bibfield
  {journal} {\bibinfo  {journal} {Quantum Science and Technology}\ }\textbf
  {\bibinfo {volume} {8}},\ \bibinfo {pages} {035002} (\bibinfo {year}
  {2023})}\BibitemShut {NoStop}%
\bibitem [{\citenamefont {Herr}\ \emph {et~al.}(2021)\citenamefont {Herr},
  \citenamefont {Obert},\ and\ \citenamefont {Rosenkranz}}]{Herr2021}%
  \BibitemOpen
  \bibfield  {author} {\bibinfo {author} {\bibfnamefont {D.}~\bibnamefont
  {Herr}}, \bibinfo {author} {\bibfnamefont {B.}~\bibnamefont {Obert}},\ and\
  \bibinfo {author} {\bibfnamefont {M.}~\bibnamefont {Rosenkranz}},\ }\bibfield
   {title} {\bibinfo {title} {Anomaly detection with variational quantum
  generative adversarial networks},\ }\href@noop {} {\bibfield  {journal}
  {\bibinfo  {journal} {Quantum Science and Technology}\ }\textbf {\bibinfo
  {volume} {6}},\ \bibinfo {pages} {045004} (\bibinfo {year}
  {2021})}\BibitemShut {NoStop}%
\bibitem [{\citenamefont {Lindauer}(2020)}]{CERT}%
  \BibitemOpen
  \bibfield  {author} {\bibinfo {author} {\bibfnamefont {B.}~\bibnamefont
  {Lindauer}},\ }\href {https://doi.org/10.1184/R1/12841247.v1} {\bibinfo
  {title} {Insider threat test dataset}} (\bibinfo {year} {2020})\BibitemShut
  {NoStop}%
\bibitem [{\citenamefont {Bergholm}\ \emph {et~al.}(2018)\citenamefont
  {Bergholm}, \citenamefont {Izaac}, \citenamefont {Schuld},\ and\
  \citenamefont {et~al}}]{Bergholm2018PennyLane}%
  \BibitemOpen
  \bibfield  {author} {\bibinfo {author} {\bibfnamefont {V.}~\bibnamefont
  {Bergholm}}, \bibinfo {author} {\bibfnamefont {J.}~\bibnamefont {Izaac}},
  \bibinfo {author} {\bibfnamefont {M.}~\bibnamefont {Schuld}},\ and\ \bibinfo
  {author} {\bibnamefont {et~al}},\ }\bibfield  {title} {\bibinfo {title}
  {Pennylane: Automatic differentiation of hybrid quantum-classical
  computations},\ }\href@noop {} {\bibfield  {journal} {\bibinfo  {journal}
  {arXiv:1811.04968}\ } (\bibinfo {year} {2018})}\BibitemShut {NoStop}%
\bibitem [{\citenamefont {Malhotra}\ \emph {et~al.}(2016)\citenamefont
  {Malhotra}, \citenamefont {Ramakrishnan}, \citenamefont {Anand},
  \citenamefont {Vig}, \citenamefont {Agarwal},\ and\ \citenamefont
  {Shroff}}]{Malhotra2016}%
  \BibitemOpen
  \bibfield  {author} {\bibinfo {author} {\bibfnamefont {P.}~\bibnamefont
  {Malhotra}}, \bibinfo {author} {\bibfnamefont {A.}~\bibnamefont
  {Ramakrishnan}}, \bibinfo {author} {\bibfnamefont {G.}~\bibnamefont {Anand}},
  \bibinfo {author} {\bibfnamefont {L.}~\bibnamefont {Vig}}, \bibinfo {author}
  {\bibfnamefont {P.}~\bibnamefont {Agarwal}},\ and\ \bibinfo {author}
  {\bibfnamefont {G.~M.}\ \bibnamefont {Shroff}},\ }\bibfield  {title}
  {\bibinfo {title} {Lstm-based encoder-decoder for multi-sensor anomaly
  detection},\ }\href@noop {} {\bibfield  {journal} {\bibinfo  {journal}
  {ArXiv}\ }\textbf {\bibinfo {volume} {abs/1607.00148}} (\bibinfo {year}
  {2016})}\BibitemShut {NoStop}%
\bibitem [{\citenamefont {Zoufal}\ \emph {et~al.}(2019)\citenamefont {Zoufal},
  \citenamefont {Lucchi},\ and\ \citenamefont {Woerner}}]{Zoufal2019}%
  \BibitemOpen
  \bibfield  {author} {\bibinfo {author} {\bibfnamefont {C.}~\bibnamefont
  {Zoufal}}, \bibinfo {author} {\bibfnamefont {A.}~\bibnamefont {Lucchi}},\
  and\ \bibinfo {author} {\bibfnamefont {S.}~\bibnamefont {Woerner}},\
  }\bibfield  {title} {\bibinfo {title} {Quantum generative adversarial
  networks for learning and loading random distributions},\ }\href@noop {}
  {\bibfield  {journal} {\bibinfo  {journal} {npj Quantum Information}\
  }\textbf {\bibinfo {volume} {5}},\ \bibinfo {pages} {1} (\bibinfo {year}
  {2019})}\BibitemShut {NoStop}%
\bibitem [{\citenamefont {Arute}\ \emph {et~al.}(2019)\citenamefont {Arute},
  \citenamefont {Arya}, \citenamefont {Babbush}, \citenamefont {Bacon},
  \citenamefont {Bardin}, \citenamefont {Barends}, \citenamefont {Biswas},
  \citenamefont {Boixo}, \citenamefont {Brandao}, \citenamefont {Buell} \emph
  {et~al.}}]{Arute2019}%
  \BibitemOpen
  \bibfield  {author} {\bibinfo {author} {\bibfnamefont {F.}~\bibnamefont
  {Arute}}, \bibinfo {author} {\bibfnamefont {K.}~\bibnamefont {Arya}},
  \bibinfo {author} {\bibfnamefont {R.}~\bibnamefont {Babbush}}, \bibinfo
  {author} {\bibfnamefont {D.}~\bibnamefont {Bacon}}, \bibinfo {author}
  {\bibfnamefont {J.~C.}\ \bibnamefont {Bardin}}, \bibinfo {author}
  {\bibfnamefont {R.}~\bibnamefont {Barends}}, \bibinfo {author} {\bibfnamefont
  {R.}~\bibnamefont {Biswas}}, \bibinfo {author} {\bibfnamefont
  {S.}~\bibnamefont {Boixo}}, \bibinfo {author} {\bibfnamefont {F.~G.}\
  \bibnamefont {Brandao}}, \bibinfo {author} {\bibfnamefont {D.~A.}\
  \bibnamefont {Buell}}, \emph {et~al.},\ }\bibfield  {title} {\bibinfo {title}
  {Quantum supremacy using a programmable superconducting processor},\
  }\href@noop {} {\bibfield  {journal} {\bibinfo  {journal} {Nature}\ }\textbf
  {\bibinfo {volume} {574}},\ \bibinfo {pages} {505} (\bibinfo {year}
  {2019})}\BibitemShut {NoStop}%
\bibitem [{\citenamefont {Zhong}\ \emph {et~al.}(2020)\citenamefont {Zhong},
  \citenamefont {Wang}, \citenamefont {Deng}, \citenamefont {Chen},
  \citenamefont {Peng}, \citenamefont {Luo}, \citenamefont {Qin}, \citenamefont
  {Wu}, \citenamefont {Ding}, \citenamefont {Hu} \emph {et~al.}}]{Zhong2020}%
  \BibitemOpen
  \bibfield  {author} {\bibinfo {author} {\bibfnamefont {H.-S.}\ \bibnamefont
  {Zhong}}, \bibinfo {author} {\bibfnamefont {H.}~\bibnamefont {Wang}},
  \bibinfo {author} {\bibfnamefont {Y.-H.}\ \bibnamefont {Deng}}, \bibinfo
  {author} {\bibfnamefont {M.-C.}\ \bibnamefont {Chen}}, \bibinfo {author}
  {\bibfnamefont {L.-C.}\ \bibnamefont {Peng}}, \bibinfo {author}
  {\bibfnamefont {Y.-H.}\ \bibnamefont {Luo}}, \bibinfo {author} {\bibfnamefont
  {J.}~\bibnamefont {Qin}}, \bibinfo {author} {\bibfnamefont {D.}~\bibnamefont
  {Wu}}, \bibinfo {author} {\bibfnamefont {X.}~\bibnamefont {Ding}}, \bibinfo
  {author} {\bibfnamefont {Y.}~\bibnamefont {Hu}}, \emph {et~al.},\ }\bibfield
  {title} {\bibinfo {title} {Quantum computational advantage using photons},\
  }\href@noop {} {\bibfield  {journal} {\bibinfo  {journal} {Science}\ }\textbf
  {\bibinfo {volume} {370}},\ \bibinfo {pages} {1460} (\bibinfo {year}
  {2020})}\BibitemShut {NoStop}%
\bibitem [{\citenamefont {Biamonte}\ \emph {et~al.}(2017)\citenamefont
  {Biamonte}, \citenamefont {Wittek}, \citenamefont {Pancotti}, \citenamefont
  {Rebentrost}, \citenamefont {Wiebe},\ and\ \citenamefont
  {Lloyd}}]{Biamonte2017}%
  \BibitemOpen
  \bibfield  {author} {\bibinfo {author} {\bibfnamefont {J.}~\bibnamefont
  {Biamonte}}, \bibinfo {author} {\bibfnamefont {P.}~\bibnamefont {Wittek}},
  \bibinfo {author} {\bibfnamefont {N.}~\bibnamefont {Pancotti}}, \bibinfo
  {author} {\bibfnamefont {P.}~\bibnamefont {Rebentrost}}, \bibinfo {author}
  {\bibfnamefont {N.}~\bibnamefont {Wiebe}},\ and\ \bibinfo {author}
  {\bibfnamefont {S.}~\bibnamefont {Lloyd}},\ }\bibfield  {title} {\bibinfo
  {title} {Quantum machine learning},\ }\href@noop {} {\bibfield  {journal}
  {\bibinfo  {journal} {Nature}\ }\textbf {\bibinfo {volume} {549}},\ \bibinfo
  {pages} {195} (\bibinfo {year} {2017})}\BibitemShut {NoStop}%
\bibitem [{\citenamefont {Mart$\acute{i}$n-Guerrero}\ and\ \citenamefont
  {Lamata}(2022)}]{Jose2022}%
  \BibitemOpen
  \bibfield  {author} {\bibinfo {author} {\bibfnamefont {J.~D.}\ \bibnamefont
  {Mart$\acute{i}$n-Guerrero}}\ and\ \bibinfo {author} {\bibfnamefont
  {L.}~\bibnamefont {Lamata}},\ }\bibfield  {title} {\bibinfo {title} {Quantum
  machine learning: A tutorial},\ }\href@noop {} {\bibfield  {journal}
  {\bibinfo  {journal} {Neurocomputing}\ }\textbf {\bibinfo {volume} {470}},\
  \bibinfo {pages} {457} (\bibinfo {year} {2022})}\BibitemShut {NoStop}%
\bibitem [{\citenamefont {Donahue}\ \emph {et~al.}(2017)\citenamefont
  {Donahue}, \citenamefont {Kr{\"a}henb{\"u}hl},\ and\ \citenamefont
  {Darrell}}]{Donahue2017AdversarialFL}%
  \BibitemOpen
  \bibfield  {author} {\bibinfo {author} {\bibfnamefont {J.}~\bibnamefont
  {Donahue}}, \bibinfo {author} {\bibfnamefont {P.}~\bibnamefont
  {Kr{\"a}henb{\"u}hl}},\ and\ \bibinfo {author} {\bibfnamefont
  {T.}~\bibnamefont {Darrell}},\ }\bibfield  {title} {\bibinfo {title}
  {Adversarial feature learning},\ }\href@noop {} {\bibfield  {journal}
  {\bibinfo  {journal} {ArXiv}\ }\textbf {\bibinfo {volume} {abs/1605.09782}}
  (\bibinfo {year} {2017})}\BibitemShut {NoStop}%
\bibitem [{\citenamefont {Bharti}\ \emph {et~al.}(2022)\citenamefont {Bharti},
  \citenamefont {Cervera-Lierta}, \citenamefont {Kyaw}, \citenamefont {Haug},
  \citenamefont {Alperin-Lea}, \citenamefont {Anand}, \citenamefont {Degroote},
  \citenamefont {Heimonen}, \citenamefont {Kottmann}, \citenamefont {Menke},
  \citenamefont {Mok}, \citenamefont {Sim}, \citenamefont {Kwek},\ and\
  \citenamefont {Aspuru-Guzik}}]{Bharti2022}%
  \BibitemOpen
  \bibfield  {author} {\bibinfo {author} {\bibfnamefont {K.}~\bibnamefont
  {Bharti}}, \bibinfo {author} {\bibfnamefont {A.}~\bibnamefont
  {Cervera-Lierta}}, \bibinfo {author} {\bibfnamefont {T.~H.}\ \bibnamefont
  {Kyaw}}, \bibinfo {author} {\bibfnamefont {T.}~\bibnamefont {Haug}}, \bibinfo
  {author} {\bibfnamefont {S.}~\bibnamefont {Alperin-Lea}}, \bibinfo {author}
  {\bibfnamefont {A.}~\bibnamefont {Anand}}, \bibinfo {author} {\bibfnamefont
  {M.}~\bibnamefont {Degroote}}, \bibinfo {author} {\bibfnamefont
  {H.}~\bibnamefont {Heimonen}}, \bibinfo {author} {\bibfnamefont {J.~S.}\
  \bibnamefont {Kottmann}}, \bibinfo {author} {\bibfnamefont {T.}~\bibnamefont
  {Menke}}, \bibinfo {author} {\bibfnamefont {W.-K.}\ \bibnamefont {Mok}},
  \bibinfo {author} {\bibfnamefont {S.}~\bibnamefont {Sim}}, \bibinfo {author}
  {\bibfnamefont {L.-C.}\ \bibnamefont {Kwek}},\ and\ \bibinfo {author}
  {\bibfnamefont {A.}~\bibnamefont {Aspuru-Guzik}},\ }\bibfield  {title}
  {\bibinfo {title} {Noisy intermediate-scale quantum algorithms},\ }\href@noop
  {} {\bibfield  {journal} {\bibinfo  {journal} {Rev. Mod. Phys.}\ }\textbf
  {\bibinfo {volume} {94}},\ \bibinfo {pages} {015004} (\bibinfo {year}
  {2022})}\BibitemShut {NoStop}%
\bibitem [{\citenamefont {Mirza}\ and\ \citenamefont
  {Osindero}(2014)}]{Mirza2014}%
  \BibitemOpen
  \bibfield  {author} {\bibinfo {author} {\bibfnamefont {M.}~\bibnamefont
  {Mirza}}\ and\ \bibinfo {author} {\bibfnamefont {S.}~\bibnamefont
  {Osindero}},\ }\bibfield  {title} {\bibinfo {title} {Conditional generative
  adversarial nets},\ }\href@noop {} {\bibfield  {journal} {\bibinfo  {journal}
  {ArXiv}\ }\textbf {\bibinfo {volume} {abs/1411.1784}} (\bibinfo {year}
  {2014})}\BibitemShut {NoStop}%
\bibitem [{\citenamefont {Benedetti}\ \emph
  {et~al.}(2019{\natexlab{b}})\citenamefont {Benedetti}, \citenamefont
  {Garcia-Pintos}, \citenamefont {Perdomo}, \citenamefont {Leyton-Ortega},
  \citenamefont {Nam},\ and\ \citenamefont {Perdomo-Ortiz}}]{Marcello2019}%
  \BibitemOpen
  \bibfield  {author} {\bibinfo {author} {\bibfnamefont {M.}~\bibnamefont
  {Benedetti}}, \bibinfo {author} {\bibfnamefont {D.}~\bibnamefont
  {Garcia-Pintos}}, \bibinfo {author} {\bibfnamefont {O.}~\bibnamefont
  {Perdomo}}, \bibinfo {author} {\bibfnamefont {V.}~\bibnamefont
  {Leyton-Ortega}}, \bibinfo {author} {\bibfnamefont {Y.}~\bibnamefont {Nam}},\
  and\ \bibinfo {author} {\bibfnamefont {A.}~\bibnamefont {Perdomo-Ortiz}},\
  }\bibfield  {title} {\bibinfo {title} {A generative modeling approach for
  benchmarking and training shallow quantum circuits},\ }\href@noop {}
  {\bibfield  {journal} {\bibinfo  {journal} {npj Quantum Information}\
  }\textbf {\bibinfo {volume} {5}},\ \bibinfo {pages} {45} (\bibinfo {year}
  {2019}{\natexlab{b}})}\BibitemShut {NoStop}%
\bibitem [{\citenamefont {Benedetti}\ \emph
  {et~al.}(2019{\natexlab{c}})\citenamefont {Benedetti}, \citenamefont {Grant},
  \citenamefont {Wossnig},\ and\ \citenamefont {Severini}}]{Benedetti2019}%
  \BibitemOpen
  \bibfield  {author} {\bibinfo {author} {\bibfnamefont {M.}~\bibnamefont
  {Benedetti}}, \bibinfo {author} {\bibfnamefont {E.}~\bibnamefont {Grant}},
  \bibinfo {author} {\bibfnamefont {L.}~\bibnamefont {Wossnig}},\ and\ \bibinfo
  {author} {\bibfnamefont {S.}~\bibnamefont {Severini}},\ }\bibfield  {title}
  {\bibinfo {title} {Adversarial quantum circuit learning for pure state
  approximation},\ }\href@noop {} {\bibfield  {journal} {\bibinfo  {journal}
  {New Journal of Physics}\ }\textbf {\bibinfo {volume} {21}},\ \bibinfo
  {pages} {043023} (\bibinfo {year} {2019}{\natexlab{c}})}\BibitemShut
  {NoStop}%
\bibitem [{\citenamefont {Cerezo}\ \emph {et~al.}(2021)\citenamefont {Cerezo},
  \citenamefont {Arrasmith}, \citenamefont {Babbush}, \citenamefont {Benjamin},
  \citenamefont {Endo}, \citenamefont {Fujii}, \citenamefont {McClean},
  \citenamefont {Mitarai}, \citenamefont {Yuan}, \citenamefont {Cincio},\ and\
  \citenamefont {Coles}}]{Cerezo2021}%
  \BibitemOpen
  \bibfield  {author} {\bibinfo {author} {\bibfnamefont {M.}~\bibnamefont
  {Cerezo}}, \bibinfo {author} {\bibfnamefont {A.}~\bibnamefont {Arrasmith}},
  \bibinfo {author} {\bibfnamefont {R.}~\bibnamefont {Babbush}}, \bibinfo
  {author} {\bibfnamefont {S.~C.}\ \bibnamefont {Benjamin}}, \bibinfo {author}
  {\bibfnamefont {S.}~\bibnamefont {Endo}}, \bibinfo {author} {\bibfnamefont
  {K.}~\bibnamefont {Fujii}}, \bibinfo {author} {\bibfnamefont {J.~R.}\
  \bibnamefont {McClean}}, \bibinfo {author} {\bibfnamefont {K.}~\bibnamefont
  {Mitarai}}, \bibinfo {author} {\bibfnamefont {X.}~\bibnamefont {Yuan}},
  \bibinfo {author} {\bibfnamefont {L.}~\bibnamefont {Cincio}},\ and\ \bibinfo
  {author} {\bibfnamefont {P.~J.}\ \bibnamefont {Coles}},\ }\bibfield  {title}
  {\bibinfo {title} {Variational quantum algorithms},\ }\href@noop {}
  {\bibfield  {journal} {\bibinfo  {journal} {Nature Reviews Physics}\ }\textbf
  {\bibinfo {volume} {3}},\ \bibinfo {pages} {625} (\bibinfo {year}
  {2021})}\BibitemShut {NoStop}%
\end{thebibliography}%
\end{document}